\documentclass[%
a4paper,twocolumn,10pt,accepted=2022-09-05,
amsmath,
amssymb,
aps
]{quantumarticle}
\pdfoutput=1
\usepackage[url=false,doi = false,isbn=false,sorting=none, style=phys]{biblatex}
\renewbibmacro{in:}{}
\usepackage{mathtools}
\usepackage{physics}
\usepackage{xcolor}
\usepackage{amsmath}
\usepackage[T1]{fontenc}
\usepackage[caption=false]{subfig} 
\usepackage{ragged2e} 
\DeclareCaptionJustification{justified}{\justifying}
\usepackage{graphicx}
\usepackage{amsthm}
\usepackage{dcolumn}
\usepackage{bm}
\usepackage[normalem]{ulem}
\usepackage[colorlinks=true, hyperindex, breaklinks, linkcolor=blue, urlcolor=blue, citecolor=blue]{hyperref}
\DeclareCaptionFormat{custom}

\newcommand{\Hil}[1]{\mathcal{H}_{#1}}
\newcommand{\Den}[1]{\mathcal{D}\left(\Hil{#1}\right)}
\newcommand{\Lin}[1]{\mathcal{L}\left(\Hil{#1}\right)}
\newcommand{\sumzeroinfty}[1]{\sum_{#1=0}^{\infty}}
\newcommand{\Cha}[3]{\mathcal{#1}_{#2\to#3}}
\newcommand{\ChaComp}[3]{\mathcal{#1}^c_{#2\to#3}}
\newcommand{\NewTr}[2]{\mathrm{Tr}_{#1}\left(#2\right)}
\newcommand{\ChaTen}[4]{\left(\mathcal{#1}\otimes\mathcal{#2}\right)_{#3\to#4}}
\newcommand{\deph}[1]{\mathcal{N}_D\left[#1\right]}
\newcommand{\loss}[1]{\mathcal{N}_L\left[#1\right]}
\newcommand{\dephcomp}[1]{\mathcal{N}^c_D\left[#1\right]}
\newcommand{\losscomp}[1]{\mathcal{N}^c_L\left[#1\right]}
\newcommand{\lossdeph}[2]{\mathcal{N}_{LD}\left[#1,#2\right]}
\newcommand{\lossdephcomp}[2]{\mathcal{N}_{LD}^c\left[#1,#2\right]}
\newcommand{\diff}{\mathrm{d}}
\newcommand{\ddiff}[2]{\frac{\diff\left(#1\right)}{\diff#2}}
\makeatletter
\newcommand*{\defeq}{\mathrel{\rlap{%
                     \raisebox{0.3ex}{$\m@th\cdot$}}%
                     \raisebox{-0.3ex}{$\m@th\cdot$}}%
                     =}
\makeatother
\newcommand{\h}[1]{\hat{#1}}

\newcommand{\kphi}{\kappa_{\phi}}
\newcommand{\kap}{\kappa}
\newcommand{\gphi}{\gamma_{\phi}}
\newcommand{\hrho}{\h{\rho}}
\newcommand{\htau}{\h{\tau}}
\newcommand{\g}{\gamma}
\newcommand{\nbar}{\bar{n}}
\newcommand{\cre}[1]{\h{#1}^{\dag}}
\newcommand{\ann}[1]{\h{#1}}
\newcommand{\num}[1]{\cre{#1}\ann{#1}}
\newcommand{\n}{\h{n}}
\newcommand{\Newmel}[3]{\langle#1\!\mid\!#2\!\mid\!#3\rangle}

\newcommand{\Newket}[1]{\lvert#1\rangle}
\newcommand{\Newbraket}[2]{\langle#1\mid#2\rangle}
\newcommand{\Newketbra}[2]{\lvert#1\rangle\langle#2\rvert}
\newcommand{\brackets}[1]{\left(#1\right)}
\newcommand{\squarebrackets}[1]{\left[#1\right]}
\newcommand{\curlbrackets}[1]{\{#1\}}
\newcommand{\Newasin}[1]{\sin^{-1}\brackets{#1}}

\newtheorem*{theorem}{Theorem}
\newtheorem{lemma}{Lemma}

\newtheorem{claim}{Claim}

\expandafter\ifcsname urlstyle\endcsname
  \else
  \fi

\begin{document}


\title{Quantum capacity and codes for the bosonic loss-dephasing channel}
\author{Peter Leviant}
\affiliation{Department of Condensed Matter Physics, Weizmann Institute of Science, Rehovot 76100, Israel}
\author{Qian Xu}
\affiliation{Pritzker School of Molecular Engineering, University of Chicago, Chicago, Illinois 60637, USA}
\author{Liang Jiang}
\affiliation{Pritzker School of Molecular Engineering, University of Chicago, Chicago, Illinois 60637, USA}
\author{Serge Rosenblum}
\email{serge.rosenblum@weizmann.ac.il}
\affiliation{Department of Condensed Matter Physics, Weizmann Institute of Science, Rehovot 76100, Israel}


\begin{abstract}
Bosonic qubits encoded in continuous-variable systems provide a promising alternative to two-level qubits for quantum computation and communication. So far, photon loss has been the dominant source of errors in bosonic qubits, but the significant reduction of photon loss in recent bosonic qubit experiments suggests that dephasing errors should also be considered. However, a detailed understanding of the combined photon loss and dephasing channel is lacking. Here, we show that, unlike its constituent parts, the combined loss-dephasing channel is non-degradable, pointing towards a richer structure of this channel. We provide bounds for the capacity of the loss-dephasing channel and use numerical optimization to find optimal single-mode codes for a wide range of error rates.
\end{abstract}

\maketitle

\section{\label{sec:introduction}Introduction} 
The presence of noise in quantum systems is one of the main challenges towards realizing useful quantum computation and communication. Indeed, qubit errors severely limit the number of gates that a quantum processor can run or the bandwidth at which a quantum communication channel can operate.

The primary method at our disposal for dealing with noise in quantum systems, apart from improving the underlying hardware, is quantum error correction \cite{Shor1995}. This requires encoding the logical information redundantly, making it resistant to the noise in the system. As such, the optimal encoding depends heavily on the noise structure, with the minimal overhead of physical modes to reliably encode a single logical qubit (or the maximal rate of communication) being determined by the quantum capacity of the noise channel \cite{Wilde2013,Lloyd1997CapacityChannel}. 

Bosonic quantum error correction uses the infinite number of levels in a single harmonic oscillator to encode quantum information redundantly. This provides a hardware-efficient alternative to traditional schemes, which encode logical qubits in arrays of physical qubits. In addition, the use of a single element substantially simplifies the noise model \cite{Ofek2016,Albert2018,Noh2020}. The most common noise types in bosonic systems are bosonic excitation loss (photon or phonon loss) and bosonic dephasing. Bosonic loss results from energy exchange with a cold environment, whereas bosonic dephasing is often caused by frequency fluctuations due to dispersive interactions with unaccounted for degrees of freedom. Another noise mechanism in bosonic systems is thermal excitation noise \cite{KyungjooNoh2020}. However, this error is less prevalent and will not be considered in this work.

Bosonic loss has often been a dominant source of noise in experimental systems. This noise channel is well understood, both in terms of its quantum capacity and the study of specific codes that protect against loss, such as the Gottesman-Kitaev-Preskill (GKP) code \cite{Albert2018,Noh2020,Gottesman2001, Campagne-Ibarcq2020}. However, recent experiments with superconducting cavities indicate that photon loss rates can be substantially reduced \cite{Romanenko2020}. Meanwhile, various physical processes (e.g., thermal excitations of coupled superconducting qubits \cite{Reagor2016QuantumQED,Rosenblum2018} and stray microwave photons \cite{Sears2012PhotonQED})   may lead to dephasing of the superconducting cavities. Hence, we expect that both loss and dephasing errors will become practically relevant errors requiring active quantum error correction. 
However, since photon number and phase are complementary observables, a tension exists between the ability to correct both error types  \cite{Grimsmo2020, Ouyang2021Trade-OffsCodes}.
This suggests a nontrivial error structure, necessitating a thorough study of the joint loss-dephasing channel.\\\\
\indent In this work, we study the bosonic loss-dephasing channel from two complementary perspectives -- through the lens of quantum information theory and quantum capacity (section \ref{sec:capacity}) and through the study of numerically optimized single-mode codes (section \ref{optimal}). We show that, unlike pure dephasing or pure loss, the combined loss-dephasing channel is \emph{not} degradable \cite{Wilde2013}. This suggests that the combined channel has a more complex structure than its constituent parts, complicating the derivation of its quantum capacity \cite{Leditzky2018DephrasureInformation}. Nonetheless, we explore the theoretical limits of the loss-dephasing channel by providing upper and lower bounds on its quantum capacity. In the second effort, we use a biconvex optimization scheme \cite{Kosut2009QuantumOptimization} to find single-mode encodings that produce an optimal encoding-decoding entanglement fidelity. Earlier work \cite{Noh2019} showed that applying this method to pure loss leads to codes similar to the hexagonal GKP code. Our results indicate that some of the optimal codes for the loss-dephasing channel are close to the numerical codes presented in Ref. \cite{Michael2016NewMode}. We also find that some optimal codes possess rotational symmetry. In the pure-loss and pure-dephasing channels, the energy constraint in the optimization problem is saturated. In contrast, the energy constraint is not typically saturated for the combined channel. Next, we compare the performance of the optimized codes to other known codes, such as GKP codes and cat codes \cite{Mirrahimi2014,Ofek2016}. Finally, we combine the two perspectives by comparing the hashing bound of the numerically optimized codes with the quantum capacity bounds (section \ref{comparison_and_hashing}).

\section{\label{sec:prelim}Preliminaries} In this section, we provide definitions for quantum channels following the notation of Ref. \cite{KyungjooNoh2020}.

Let $\Hil{X}$ be the Hilbert space of a system $X$, $\Lin{X}$ the space of linear operators over $\Hil{X}$ \cite{Wilde2013}, and $\Den{X}$ the space of density matrices. A linear map $\Cha{N}{A}{B}\colon \Lin{A}\to\Lin{B}$ is called a \emph{quantum channel} from a sender $A$ to a receiver $B$ if it is completely positive and trace-preserving (CPTP), or equivalently, if for any system $W$, the map 
\begin{equation}
\begin{split}
    &\ChaTen{N}{I}{AW}{BW}\colon\Lin{A}\otimes\Lin{W}\\&\to\Lin{B}\otimes\Lin{W},
\end{split}
\end{equation} defined by $\hrho_A\otimes\htau_W\mapsto\mathcal{N}(\hrho_A)\otimes\htau_W$, maps density matrices to density matrices.

Any quantum channel $\Cha{N}{A}{B}$ can be written as a sum of \emph{Kraus operators}, which are linear maps $K_i\colon\Hil{A}\to\Hil{B}$ such that \begin{equation}
    \Cha{N}{A}{B}=\sum_{i}K_i\bullet K_i^{\dag},
\end{equation}
where $\bullet$ is any operator in $\Lin{A}$.
\noindent Alternatively, every quantum channel $\Cha{N}{A}{B}$ can be represented using an isometric extension $U\colon \Hil{A}\to\Hil{B}\otimes\Hil{E}$ with $E$ denoting the environment. The channel from $A$ to $B$ is then obtained by tracing out the environment: 
\begin{equation}
    \Cha{N}{A}{B}=\NewTr{E}{\Cha{U}{A}{BE}},\,\Cha{U}{A}{BE}\defeq U\bullet U^{\dag}.
\end{equation}
This isometric extension is called a \emph{Stinespring dilation}, and it is unique up to an isometry on the environment.

The channel obtained by tracing over $B$ is called the \emph{complementary channel} of $\Cha{N}{A}{B}$ and is denoted by $\ChaComp{N}{A}{E}=\NewTr{B}{\Cha{U}{A}{BE}}$.

Finally, a channel is said to be \emph{degradable} (\emph{anti-degradable}) if the receiver (environment) can simulate the information obtained by the environment (the receiver). Formally, the channel $\Cha{N}{A}{B}$ is degradable (anti-degradable) if there exists a \emph{degradation} (\emph{anti-degradation}) \emph{channel} $\Cha{D}{B}{E}$ ( $\Cha{D}{E}{B}^{\prime}$) such that $\mathcal{N}^c_{A\to E}=\mathcal{D}_{B\to E}\circ\mathcal{N}_{A\to B}$ ($\mathcal{N}_{A\to B}=\mathcal{D}^{\prime}_{E\to B}\circ\mathcal{N}^c_{A\to E}$), where $\circ$ denotes channel composition. The property of  degradability (anti-degradability) is invariant to the specific choice of Stinespring dilation. 

\subsection{\label{sec:dephasing}Bosonic dephasing channel}
We define the \emph{bosonic pure-dephasing channel} as:
\begin{equation}
\begin{split}
    &\deph{\gphi}\brackets{\hrho} =\\ &\sumzeroinfty{m,n}e^{-\frac{1}{2}\gphi(m-n)^2}\Newmel{m}{\hrho}{n}\Newketbra{m}{n},\,\gphi\geq0,
\end{split}
\end{equation}
where $\gamma_{\phi}$ characterizes the dephasing strength and $|n\rangle$ is the Fock state with $n$ excitations.

There are multiple other ways to represent the dephasing channel. It can be described by the Lindblad master equation

\begin{equation}
\ddiff{\hrho(t)}{t}=\kphi\mathcal{D}\squarebrackets{\n}\brackets{\hrho(t)},    
\end{equation}
where $\kphi$ is the dephasing rate ($\gphi=\kphi t$), $\hrho(t)\defeq\deph{\kphi t}\brackets{\hrho(0)}$ is the state evolution,  $\mathcal{D}[\h{A}]\defeq\ann{A}\bullet\cre{A}-\frac{1}{2}\{\num{A},\bullet\}$ is the Lindbladian with jump operator $\h{A}$, and $\n=\num{a}$ is the number operator of the bosonic mode with annihilation operator $\hat{a}$.

The dephasing channel can also be represented as an integral over continuously many Kraus operators, each representing a random phase rotation:

\begin{equation}
\deph{\gphi}(\hrho)=\frac{1}{\sqrt{2\pi\gphi}}\int_{-\infty}^{\infty}e^{-\frac{\phi^2}{2\gphi}}e^{i\phi\n}\hrho e^{-i\phi\n}\diff\phi,
\end{equation}
or as a sum over a discrete set of Kraus operators:
\begin{equation}
\deph{\gphi}=\sumzeroinfty{k}\ann{D}_k\bullet\cre{D}_k,\,\ann{D}_k=\sqrt{\frac{\gphi^k}{k!}}e^{-\frac{\gphi}{2}\n^2}\n^k.
\end{equation}
These two Kraus representations correspond to different Stinespring dilations. In particular, the discrete representation is due to a \emph{conditional displacement} evolution \cite{Arqand2020}:
\begin{equation}
\begin{split}
    U&=e^{\sqrt{\gphi}\num{a}(\cre{b}-\ann{b})}\\\deph{\gphi}\brackets{\hrho}&=\NewTr{E}{\ann{U}\brackets{\hrho_X\otimes\Newketbra{0}{0}_E}\cre{U}},
\end{split}
\end{equation}
where $\h{b}$ is the annihilation operator of the environment mode. In this representation, the complementary channel can be written as:
\begin{equation}
\dephcomp{\gphi}\brackets{\hrho}=\sumzeroinfty{n}\Newmel{n}{\hrho}{n}\Newketbra{\sqrt{\gphi}n}{\sqrt{\gphi}n},
\end{equation}
with $\Newket{\sqrt{\gphi}n}$ a coherent state with amplitude $\sqrt{\gphi}n$.
The dephasing channel is degradable, since  
\begin{equation}
\begin{split}
\dephcomp{\gphi} & =\mathcal{D}\circ\deph{\gphi}  \\
 \text{ with }\Cha{D}{X}{E}&=\sumzeroinfty{n}\Newketbra{\sqrt{\gphi}n}{n}\bullet\Newketbra{n}{\sqrt{\gphi}n}. \\
\end{split}    
\end{equation}

The channel $\mathcal{N}_D$ is diagonal in the operator basis $\curlbrackets{\Newketbra{m}{n}\,|\,m,n\geq0}$. This implies the following \emph{covariance property} of $\mathcal{N}_D$ with respect to any operator $\h{A}$ that is diagonal in the Fock basis: \begin{equation}
\brackets{\ann{A}\bullet\cre{A}}\circ\deph{\gphi} = \deph{\gphi}\circ \brackets{\ann{A}\bullet\cre{A}}.
\end{equation}
In particular, this identity is satisfied for phase-space rotations $\h{A}=e^{i\theta\h{n}},\,\theta\in\mathbb{R}$.

\subsection{\label{sec:loss}Bosonic loss channel} 
The bosonic loss or photon loss channel $\mathcal{N}_L[\gamma]$ can be defined using the master equation
\begin{equation}
    \loss{1-e^{-\kap t}}\brackets{\hrho(0)}=\hrho(t),\,\ddiff{\hrho(t)}{t}=\kap\mathcal{D}\squarebrackets{\ann{a}}\brackets{\hrho(t)},
\end{equation}
where $\gamma=1-e^{-\kappa t},\,\gamma\in\squarebrackets{0,1}$.
Alternatively, we can use the Kraus representation:
\begin{equation}
\loss{\g}=\sumzeroinfty{n}\ann{L}_k\bullet\cre{L}_k,\,\ann{L}_k=\sqrt{\frac{\g^k}{k!}}(1-\g)^{\frac{\h{n}}{2}}\h{a}^k.    
\end{equation}
The Stinespring dilation of $\loss{\gamma}$ takes the form of a beam-splitter interaction:
\begin{equation}
\begin{split}
    &U=  e^{\Newasin{\sqrt{\g}}(\cre{b}\ann{a}-\cre{a}\ann{b})},
    \,
    \\&\loss{\g}\brackets{\hrho}=\NewTr{E}{\ann{U}\brackets{\hrho_X\otimes\Newketbra{0}{0}_E}\cre{U}}.
\end{split}
\end{equation}
Since for all coherent states
$\Newket{\alpha}_X$
\begin{equation}
    U\Newket{\alpha,0}_{XE}=\Newket{\sqrt{1-\gamma}\alpha,\sqrt{\gamma}\alpha}\label{energy_split},
\end{equation} 
the complementary channel takes on a simple form
\begin{equation}
    \losscomp{\g}=\loss{1-\g}.
\end{equation}
Because $\loss{1-\eta_1\eta_2}=\loss{1-\eta_1}\circ\loss{1-\eta_2}$, it follows that $\loss{\g}$ is degradable (anti-degradable) for  $\gamma\leq\frac{1}{2}$ ($\gamma\geq\frac{1}{2}$).

Finally, using Eq. \ref{energy_split}, it can be shown that the bosonic loss channel is also covariant with respect to rotations $\ann{U}=e^{i\theta\h{n}}$.
\subsection{\label{sec:lossanddephasing}Joint loss and dephasing channel}
The bosonic loss-dephasing channel $\lossdeph{\g}{\gphi}$ arises from combining the previous two noise mechanisms. It can be defined using the following master equation:
\begin{equation}
    \begin{split}
        &\lossdeph{1-e^{-\kap t}}{\kphi t}\brackets{\hrho(0)}=\hrho(t)\\
        &\ddiff{\hrho(t)}{t}=\brackets{\kap\mathcal{D}\squarebrackets{\ann{a}}+\kphi\mathcal{D}\squarebrackets{\num{a}}}\brackets{\hrho(t)}.
    \end{split}
\end{equation}
This expression can be greatly simplified by noting that the two Lindbladians commute -- $\mathcal{D}\squarebrackets{\h{a}}\mathcal{D}\squarebrackets{\num{a}}=\mathcal{D}\squarebrackets{\num{a}}\mathcal{D}\squarebrackets{\h{a}}$, yielding \begin{equation}
    \lossdeph{\g}{\gphi}=\loss{\g}\circ\deph{\gphi}=\deph{\gphi}\circ\loss{\g}.
\end{equation}
We can use this identity to form a Stinespring dilation for the combined channel using two environments and the dilations of the pure-loss and pure-dephasing channels. To do so, let $\h{a},\h{b}_l,\h{b}_d$ be the annihilation operators corresponding to modes of the system $X$, a first environment $E_l$ of the loss channel and a second environment $E_d$ of the dephasing channel. The isometric extension is
\begin{equation}\label{stinespringoflossdeph}
\begin{split}
    U&= e^{\sqrt{\gphi}\num{a}(\cre{b}_d-\ann{b}_d)}e^{\Newasin{\sqrt{\g}}(\cre{b}_l\ann{a}-\cre{a}\ann{b}_l)}\\\lossdeph{\g}{\gphi}\brackets{\hrho}&=\NewTr{E_lE_d}{\ann{U}\brackets{\hrho_X\otimes\Newketbra{00}{00}_{E_lE_d}}\cre{U}}.
\end{split}
\end{equation}
Finally, the loss-dephasing channel inherits covariance with respect to rotations from its constituent channels.
\section{\label{sec:capacity}Quantum capacity}
The \emph{quantum capacity} $Q_{\mathcal{N}}$ of a channel $\mathcal{N}$ is the highest rate at which one can reliably transmit quantum information over many uses of the channel \cite{Wilde2013}. The quantum capacity is an important metric for evaluating the usability of the channel for quantum communication and storage. It can be shown \cite{Wilde2013} that $Q_{\mathcal{N}}$ is identical to the regularized maximal coherent information, which is a regularized limit over the coherent information that one can transmit using asymptotically many copies of $\mathcal{N}$:
\begin{equation}\label{eq:regularized}
\begin{split}
    &Q_{\Cha{N}{X}{Y}} = I_{c,\text{reg}}(\mathcal{N})=\\& \lim_{n\to\infty}\frac{1}{n}\squarebrackets{\max_{\hrho\in\Den{X^n}}I_c\brackets{\hrho,\mathcal{N}^{\otimes n}}},
\end{split}
\end{equation}
where $
    I_c(\hrho,\mathcal{M})\coloneqq H(\mathcal{M}(\hrho))-H(\mathcal{M}^c(\hrho))$ denotes the coherent information and $
    H(\hrho)\coloneqq -\NewTr{}{\hrho\log_2\hrho}$ is the von Neumann entropy.

Since Eq. \ref{eq:regularized} contains an optimization over all possible input states for asymptotically many blocks, it is generally difficult to calculate the quantum capacity. However, this calculation is greatly simplified for degradable and anti-degradable channels. The capacity of anti-degradable channels is zero, which can be intuitively understood from the no-cloning theorem. For degradable channels, the regularized maximal coherent information equals the single-shot coherent information maximized over all possible input states $\hrho$ \cite{Wilde2013}:
\begin{equation}
Q_{\Cha{N}{X}{Y}}=\max_{\hrho\in\Den{X}}I_c(\hrho,\mathcal{N}).    
\end{equation}
Moreover, $I_c(\hrho,\mathcal{N})$ is concave in $\hrho$, allowing its calculation using convex optimization methods. In particular, if $\mathcal{N}$ is a rotationally covariant degradable channel on a single bosonic mode (such as pure dephasing or pure loss), we have
\begin{equation}
    \begin{split}
        Q_{\mathcal{N}}=&\max_{\hrho\in\Den{X}}I_c(\hrho,\mathcal{N})\\=&\max_{\hrho\in\Den{X}}\int_0^{2\pi}\frac{\diff\theta}{2\pi} I_c(\hrho,\brackets{e^{i\theta\h{n}} e^{-i\theta\h{n}}}\circ\mathcal{N})\\=&\max_{\hrho\in\Den{X}}\int_0^{2\pi}\frac{\diff\theta}{2\pi} I_c(e^{i\theta\h{n}}\hrho e^{-i\theta\h{n}},\mathcal{N})\\\leq&\max_{\hrho\in\Den{X}}I_c(\int_0^{2\pi}\frac{\diff\theta}{2\pi} e^{i\theta\h{n}}\hrho e^{-i\theta\h{n}},\mathcal{N})\\=&\max_{\hrho\in\Den{X}}I_c(\sumzeroinfty{n}\Newmel{n}{\hrho}{n}\Newketbra{n}{n},\mathcal{N})\\=&\max_{\hrho\in\Den{X}\text{ diagonal}}I_c(\hrho,\mathcal{N})\leq Q_{\mathcal{N}},
    \end{split}
\end{equation} 
where the inequality follows from the concavity of $I_c(\bullet,\mathcal{N})$ for degradable channels. Therefore, to calculate $Q_{\mathcal{N}}$, it suffices to optimize only over diagonal states (a similar observation was made in Ref. \cite{Arqand2020}).

 When the bosonic mode has finite energy, which is of practical interest, we can define the energy-constrained channel capacity \cite{Winter2017Energy-constrainedCapacitiesb}:
\begin{equation}
\begin{split}
    &Q^{\leq \bar{n}}_{\mathcal{N}}=\\&\lim_{n\to\infty}\frac{1}{n}\squarebrackets{\max_{\hrho\in\Den{X^n},\frac{1}{n}\sum_i {\NewTr{}{\h{n}_i\hrho}}\leq \bar{n}}I_c\brackets{\hrho,\mathcal{N}^{\otimes n}}},
\end{split}
\end{equation}
where $\bar{n}$ is the maximum allowed mean occupation number per channel use.

\subsection{\label{sec:previous}Previous results} 
As mentioned earlier, pure-loss and pure-dephasing channels are degradable, making their quantum capacities relatively easy to evaluate. 

Indeed, photon loss -- a prominent form of Gaussian noise -- is well understood from a quantum information theoretic viewpoint \cite{Noh2019,Wolf2007QuantumChannels}. Its quantum capacity is given by \cite{Weedbrook2012GaussianInformation} \begin{equation}
    Q_{\loss{\g}}=\max\squarebrackets{\log_2\brackets{\frac{1-\g}{\g}},0}
\end{equation} and in the energy-constrained case \cite{Noh2019,Wilde2018Energy-constrainedChannels}, 
\begin{equation}
Q^{\leq \bar{n}}_{\loss{\g}}=\max\squarebrackets{g((1-\g)\bar{n})-g(\g \bar{n}),0},    
\end{equation}
where $g(\bar{n})=H(\htau(\bar{n}))$ is the entropy of the thermal state $\htau(\nbar)$ with mean occupation number $\bar{n}$, namely $\htau(\nbar)=\sumzeroinfty{n}\frac{\nbar^n}{(1+\nbar)^{n+1}}\Newketbra{n}{n}$. This implies that thermal states, which are diagonal in the Fock basis and have a geometric photon number distribution, achieve the upper bound on coherent information. It was also shown in Ref. \cite{Noh2019} that a multi-mode encoding using GKP states \cite{Gottesman2001} on a $2N$-dimensional lattice achieves the quantum capacity up to a constant offset of $\log_2 e$ for all $\gamma\leq \frac{1}{2}$ in the limit $N\to \infty$. 

Recently, the quantum capacity of the pure-dephasing channel was shown by Lami et al. \cite{Lami2022ExactChannels} to be equal to 
\begin{equation}
\begin{split}
    &Q_{\deph{\gphi}} = D(p||u)\\ &= \log_2\varphi(e^{-\gphi})+\frac{2}{\ln 2}\sum_{k=1}^{\infty}\frac{(-1)^{k-1}e^{-\frac{\gphi}{2}(k^2+k)}}{k(1-e^{-k\gphi})},
\end{split}
\end{equation}
where $D(p||u)$ is the relative entropy between the wrapped normal distribution $p$ and the uniform distribution $u$, and $\varphi(q)\defeq\Pi_{k=1}^{\infty}(1-q^{k})$ is the Euler function.

The energy-constrained capacity of the pure-dephasing channel was numerically analyzed by Arqand et al. \cite{Arqand2020}. 
In Appendix \ref{app:dephasing}, we show numerically that the coherent information of thermal states approximates the energy-constrained capacity of the pure-dephasing channel. 
\subsection{\label{sec:capacityofthejointchannel}Quantum capacity of the loss-dephasing channel} The quantum capacity of the loss-dephasing channel $\lossdeph{\g}{\gphi}$ is difficult to evaluate for $\g,\gphi\neq0$. Indeed, unlike its constituent parts, the combined channel is \emph{not} degradable. Similar behavior also appears in concatenated qubit channels \cite{Leditzky2018DephrasureInformation,Siddhu2021PositivityChannels}. We can nonetheless provide insight into the quantum capacity by providing upper and lower bounds.
\subsubsection{\label{nondegradability} Non-degradability of the loss-dephasing channel} 
One of the main results of this work is a proof that the loss-dephasing channel is non-degradable.
\begin{theorem}\label{nondegradabilitythm}
For $\gphi>0$ and $1\geq\g>0$, the loss-dephasing channel $\lossdeph{\g}{\gphi}$ is not degradable. For $\gamma\geq1/2$, the channel is anti-degradable.
\end{theorem}
\begin{proof}
We briefly outline the proof of the theorem. A complete proof based on four lemmas is provided in appendix \ref{sec:nondegradabilityappendix}.

According to Lemma \ref{lem:compchannel} and Lemma \ref{lem:degrading}, for $\g<1$, there is a Stinespring dilation of $\lossdeph{\g}{\gphi}$ with a unique degradation map $\mathcal{D}$ given by Eq. \ref{eq:degradation_super}. For $\g=1$, no such map exists. Since all dilations differ by an isometry on the environment, the existence of a degradation channel for one dilation implies its existence for all dilations. We prove in Lemma \ref{lem:notchannel} that the map $\mathcal{D}$ is not a quantum channel for $1>\g>0$, $\gphi>0$. This implies that $\lossdeph{\g}{\gphi}$ is not degradable for $\g,\gphi>0$. 

To prove the second part of the theorem, we rely on the fact that for $\g\geq 1/2$, the loss-dephasing channel is a concatenation of an anti-degradable channel $\loss{\g}$ with another channel. Therefore, the information lost to the first environment ($E_l$) is sufficient to simulate the entire channel, proving its anti-degradability. In more detail, since $\loss{\g}$ is anti-degradable for $\g\geq\frac{1}{2}$, there exists an anti-degradation channel $\mathcal{D}'$ such that $\loss{\g}=\mathcal{D}'\circ\losscomp{\g}$. Concatenating both sides with $\deph{\gphi}$ yields $$\lossdeph{\g}{\gphi}=\deph{\gphi}\circ\loss{\g}=\deph{\gphi}\circ\mathcal{D}'\circ\losscomp{\g}.$$ Using the Stinespring dilation for $\lossdeph{\g}{\gphi}$ from Eq. \ref{stinespringoflossdeph}, we can write the complementary channel as 
$$
\lossdephcomp{\g}{\gphi}\brackets{\hrho_X}=\NewTr{X}{\ann{U}\brackets{\hrho_X\otimes\Newketbra{00}{00}_{E_lE_d}}\cre{U}},
$$
with $U=e^{\sqrt{\gphi}\num{a}(\cre{b}_d-\ann{b}_d)}e^{\Newasin{\sqrt{\g}}(\cre{b}_l\ann{a}-\cre{a}\ann{b}_l)}$. Note that $U$ is the product of two unitaries $U=U^d U^l$, where $U^l$ acts only on $X$ and $E_l$ and $U^d$ acts only on $X$ and $E_d$, so that $\mathrm{Tr}_{E_dX}\circ\brackets{U^{d}\bullet\brackets{U^d}^{\dagger}}=\mathrm{Tr}_{E_dX}$. Therefore,
\begin{widetext}
\begin{equation*}
    \begin{split}
        \brackets{\mathrm{Tr}_{E_d}\circ\lossdephcomp{\g}{\gphi}}\brackets{\hrho_X}=&\brackets{\mathrm{Tr}_{E_dX}\circ\brackets{U^{d}\bullet\brackets{U^d}^{\dagger}}}\brackets{U^{l}\brackets{\hrho_X\otimes\Newketbra{00}{00}_{E_lE_d}}\brackets{U^l}^{\dagger}}\\=& \NewTr{E_dX}{U^l\brackets{\hrho_X\otimes\Newketbra{00}{00}_{E_lE_d}}\brackets{U^l}^{\dagger}}=\losscomp{\g}\brackets{\hrho_X},
    \end{split}
\end{equation*}
\end{widetext}
since $U^l$ is a Stinespring dilation of $\loss{\g}$. Combining the expressions above, we obtain 
$$
\lossdeph{\g}{\gphi}=\deph{\gphi}\circ\mathcal{D}'\circ\mathrm{Tr}_{E_d}\circ\lossdephcomp{\g}{\gphi},
$$ proving anti-degradability for $\g\geq\frac{1}{2}$.
\end{proof}
\subsubsection{\label{upperbound} Data processing upper bound} Since the loss-dephasing channel can be written as the composition of a loss channel and a dephasing channel, a data processing argument \cite{Wilde2013} implies that its capacity is smaller than that of each of its two constituent channels. This inequality allows the derivation of an upper bound:

\begin{equation}
    \begin{split}
        Q^{\leq \nbar}_{\lossdeph{\g}{\gphi}}&\leq Q_{\text{data processing}}^{\leq \nbar}(\g,\gphi)\\&\defeq\min\curlbrackets{Q^{\leq \nbar}_{\loss{\g}},Q^{\leq \nbar}_{\deph{\gphi}}}.
    \end{split}
    \label{eq:data_processing_bound}
\end{equation}

\subsubsection{\label{lowerbound} Single-mode lower bound} The coherent information of any single-mode input state provides a lower bound on the capacity. A good choice for a representative state is a thermal state that saturates the energy bound, since the coherent information of this state equals the capacity in the pure loss case and approximates the capacity in the pure-dephasing case (see appendix \ref{app:dephasing}). More generally, we can optimize the coherent information over all diagonal states:
\begin{widetext}
\begin{equation}
    \begin{split}
        Q^{\leq \nbar}_{\lossdeph{\g}{\gphi}}\geq& \max_{\hrho\in\Den{X},\,\NewTr{}{\hrho \h{n}}\leq \nbar}I_c(\hrho,\lossdeph{\g}{\gphi})\geq Q_{\text{diagonal}}^{\leq \nbar}(\g,\gphi)\defeq\max_{\overset{\hrho\in\Den{X}}{\NewTr{}{\hrho \h{n}}\leq \nbar,\,\hrho\text{ diagonal}}}I_c(\hrho,\lossdeph{\g}{\gphi})\\\geq& Q_{\text{thermal}}^{\nbar}(\g,\gphi)\defeq I_c(\htau(\nbar),\lossdeph{\g}{\gphi}),
    \end{split}\label{eq:lower_bound}
\end{equation}
\end{widetext}
where $Q_{\text{diagonal}}$ and $Q_{\text{thermal}}$ are lower bounds obtained by calculating the one-shot coherent information using the optimal diagonal state and the thermal state $\htau(\nbar)$, respectively.

Since the loss-dephasing channel is not degradable, the one-shot coherent information is no longer concave, and we are not guaranteed to obtain a single maximum. As a result, a numerical optimization over diagonal states might not converge to a global maximum. Furthermore, while the channel is covariant to rotations, we cannot use a concavity argument to show that a diagonal state achieves the optimal single-mode coherent information. However, as we will show later,  $Q_{\text{diagonal}}$ and $Q_{\text{thermal}}$ are tight lower bounds on the capacity for $\g\ll1$ or $\gphi\ll1$.

\begin{figure*}
  \includegraphics[width=\textwidth]{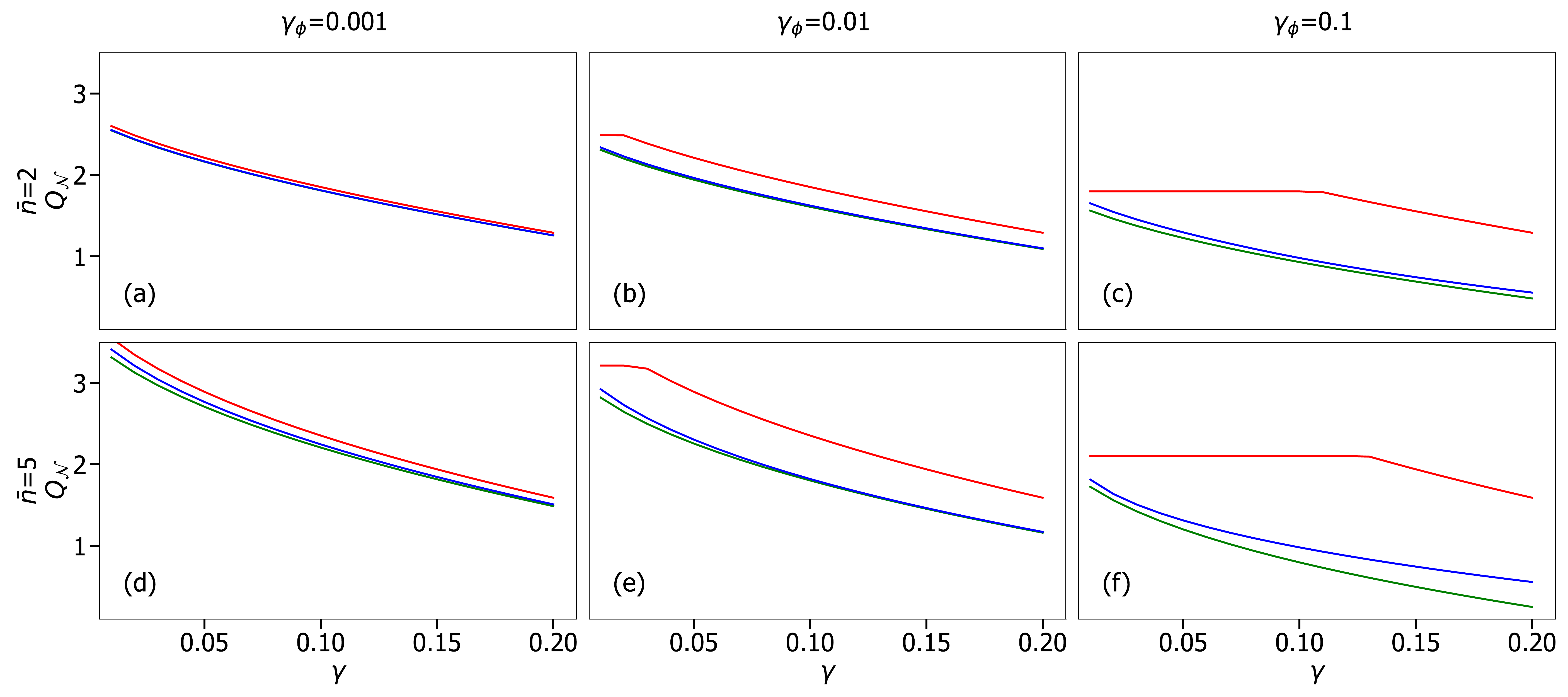}
  \caption{Bounds on the quantum capacity of the loss-dephasing channel. Panels (a) to (f) depict the bounds as a function of the loss rate $\g$ for various dephasing rates $\gphi$ (columns) and energy constraints $\bar{n}$ (rows). The red lines show the data processing upper bounds. The green lines are the lower bounds derived from the coherent information of thermal states. The blue lines represent optimizations over the coherent information of diagonal states.}
  \label{fig:capacitycomparison}
\end{figure*}
\subsubsection{\label{compbound} Comparison of bounds} Here, we compare the previously derived bounds on the quantum capacity of the loss-dephasing channel for several dephasing rates $\gphi\in\{0.001,0.01,0.1\}$, mean energies $\nbar\in\{2,5\}$, and various loss rates $\gamma\in [0.01,0.2]$. The results of this comparison are shown in Fig. \ref{fig:capacitycomparison}. The lower bound $Q_{\textrm{thermal}}$ is almost as tight as the lower bound $Q_{\textrm{diagonal}}$ obtained by optimizing over diagonal states, except for large $\gamma$ and $\gamma_{\phi}$, i.e. $\gamma, \gamma_{\phi} \gtrsim 0.1$. Both the lower bounds and the upper bounds are tight when the joint channel is dominated by either loss or dephasing, i.e., $\gamma \rightarrow 0$ or  $\gamma_{\phi} \rightarrow 0$. However, the bounds become looser when both $\gamma$ and $\gamma_{\phi}$ are large, as the joint channel deviates further from a degradable channel. Overall, both the lower and upper bounds increase as more energy is allowed for the bosonic mode. Another feature of the data processing upper bound $Q_{\textrm{data processing}}$ is that, given $\gamma_{\phi}$, $Q_{\textrm{data processing}}$ first increases as $\gamma$ decreases and then saturates when $\gamma\lesssim\gamma_{\phi}$. This feature appears because $Q_{\textrm{data processing}}$ is, by definition, the minimum of the separate pure-loss and pure-dephasing channel capacities (see Eq.~\ref{eq:data_processing_bound}), which equals the latter (as a constant) when dephasing is dominant.

\section{\label{optimal} Numerically optimized error correction codes} 

Tailored encoding and decoding operations are required to faithfully transmit quantum information over a noisy channel  $\Cha{N}{A}{B}$. We use the entanglement fidelity \cite{Wilde2013} of the composite encoding-noise-decoding channel $\Cha{E}{M_A}{M_B}=\Cha{R}{B}{M_B}\circ \Cha{N}{A}{B}\circ \Cha{S}{M_A}{A}$ ($M_A,M_B$ are identical message spaces available to $A$ and $B$, respectively) as the figure of merit characterizing how well the information is preserved through $\mathcal{E}$. The optimal encoding and decoding strategy is thus given by 

\begin{equation}
(\mathcal{S},\mathcal{R})_{\text{opt}} = \underset{(\mathcal{S},\mathcal{R})}{\text{argmax}}\,F_{\mathcal{E}},    
\end{equation}
where $F_{\mathcal{E}}$ is the entanglement fidelity of the composite channel, defined as the overlap between a maximally entangled state $|\Gamma\rangle$, and the state  $\hat{\rho}_{\mathcal{E}}$ obtained after one part of $|\Gamma\rangle$ is transmitted through $\mathcal{E}$:

\begin{equation}\label{eq:entfid}
    \begin{split}
        F_{\mathcal{E}} =& \Newmel{\Gamma}{\hrho_{\mathcal{E}}}{\Gamma}=\frac{1}{\brackets{\dim\Hil{M_A}}^2}\sum_{i,j}\Newmel{i}{\mathcal{E}(\Newketbra{i}{j})}{j}\\=\colon&\frac{1}{\brackets{\dim\Hil{M_A}}^2}\mathrm{Tr}\mathcal{E},\\
        \hrho_{\mathcal{E}}\equiv& \brackets{\mathcal{E}\otimes \mathcal{I}}\brackets{\Newketbra{\Gamma}{\Gamma}},\,\Newket{\Gamma}=\frac{1}{\sqrt{\dim{\Hil{M_A}}}}\sum_{i}\Newket{ii},
    \end{split}
\end{equation}
where $\{\Newket{i}\}$ is a basis of $M_A\cong M_B$ and $\mathrm{Tr}\mathcal{E}$ is the trace of a matrix representation of $\mathcal{E}$.

If $A$ is a bosonic system, we can modify this definition to handle energy-constrained encodings by defining 
\begin{equation}
(\mathcal{S},\mathcal{R})_{\text{opt}}^{\leq \nbar} = \underset{(\mathcal{S},\mathcal{R}),\,\NewTr{}{\n \mathcal{S}(\h{\pi}})\leq \nbar}{\text{argmax}}F_{\mathcal{E}},    
\end{equation}
where $\h{\pi}$ is a maximally mixed state in $M_A$.

Optimizing over either of the channels $\mathcal{S}$ or $\mathcal{R}$ while keeping the other fixed is a convex optimization problem. As such, this problem has a unique solution that is efficient to compute using semidefinite programming \cite{Kosut2009QuantumOptimization}. This insight leads to an intuitive biconvex optimization algorithm in which we iteratively find the optimal decoding given an updated encoding, and vice versa. However, the biconvex problem itself is not guaranteed to be convex. As a result, there might be cases where the algorithm does not converge or where the converged encoding depends on the chosen initial encoding.

Our numerical procedure optimizes the entanglement fidelity, but does not consider other important figures of merit. These include the performance of the encoding and decoding procedures themselves, or the difficulty in their experimental realization. In addition, the constraint on the average photon number fails to take into account the spread of the codes in Fock space. For example, the geometric distribution of GKP codes occupies a much larger number of energy levels than Poissonian-distributed cat codes with the same $\nbar$. Finally, increasing the entanglement fidelity of a code might not always increase its capacity to transmit information, as we will show in section \ref{hashing}.

For further details and results on the pure-loss channel, we refer to section 5 in Ref. \cite{Noh2019}. In the next section, we expand on these results by adding dephasing noise. We find that the optimization algorithm consistently converges to unique encodings for various loss and dephasing rates.

\subsection{\label{biconvexresults} Discussion of optimization results}
To gain insight into the structure of numerically optimized qubit codes, we perform the optimization for various loss and dephasing rates. Since the loss-dephasing channel is covariant under rotations, rotating a code does not alter its performance. For the pure-dephasing channel, the optimized codes are also covariant under diagonal unitaries. Therefore, we regularize the codes accordingly after the optimization process (see Appendix \ref{app:regularizationcompbound}).

As long as the error rates are sufficiently high and the energy constraint sufficiently low, we consistently converge to the same optimal codes for a given triplet $\g,\gphi,\nbar$. The optimization results are shown in Fig. \ref{fig:phaseplot}. Each plot represents an optimization result for a specific pair of loss and dephasing rates -- that is, a local maximum of the entanglement fidelity. These local maxima are most likely also global maxima, since we observe that optimization runs with different randomly chosen initial codes converge to the same optimal code. 

As we allow for higher $\nbar$ or lower $\g,\gphi$,  the optimal value of the entanglement fidelity approaches one and the landscape becomes shallower, allowing for more local maxima with similar entanglement fidelities to appear. This causes the optimization result to depend on the initial state. Such parameter ranges are not considered and are represented in Fig. \ref{fig:phaseplot} by a shaded region.

A key takeaway from these results is that while numerically optimized codes for pure-loss or pure-dephasing channels saturate the mean energy constraint, this is not generally the case. Indeed, for the combined loss-dephasing channel with $\g,\gphi>0$, the numerically optimized codes have a particular mean energy that does not vary when allowing higher energies. 

\section{\label{comparison_and_hashing} Comparison with known codes} Various quantum error-correcting codes have been previously developed to protect against bosonic noise. For a short overview of GKP codes (which protect well against loss) \cite{Gottesman2001}, rotation codes \cite{Grimsmo2020} and numerical codes \cite{Michael2016NewMode}, see appendix \ref{app:knowncodes}. Here, we show how these codes compare to the numerically optimized codes from this work. 

\begin{figure}
  \includegraphics[width=\linewidth]{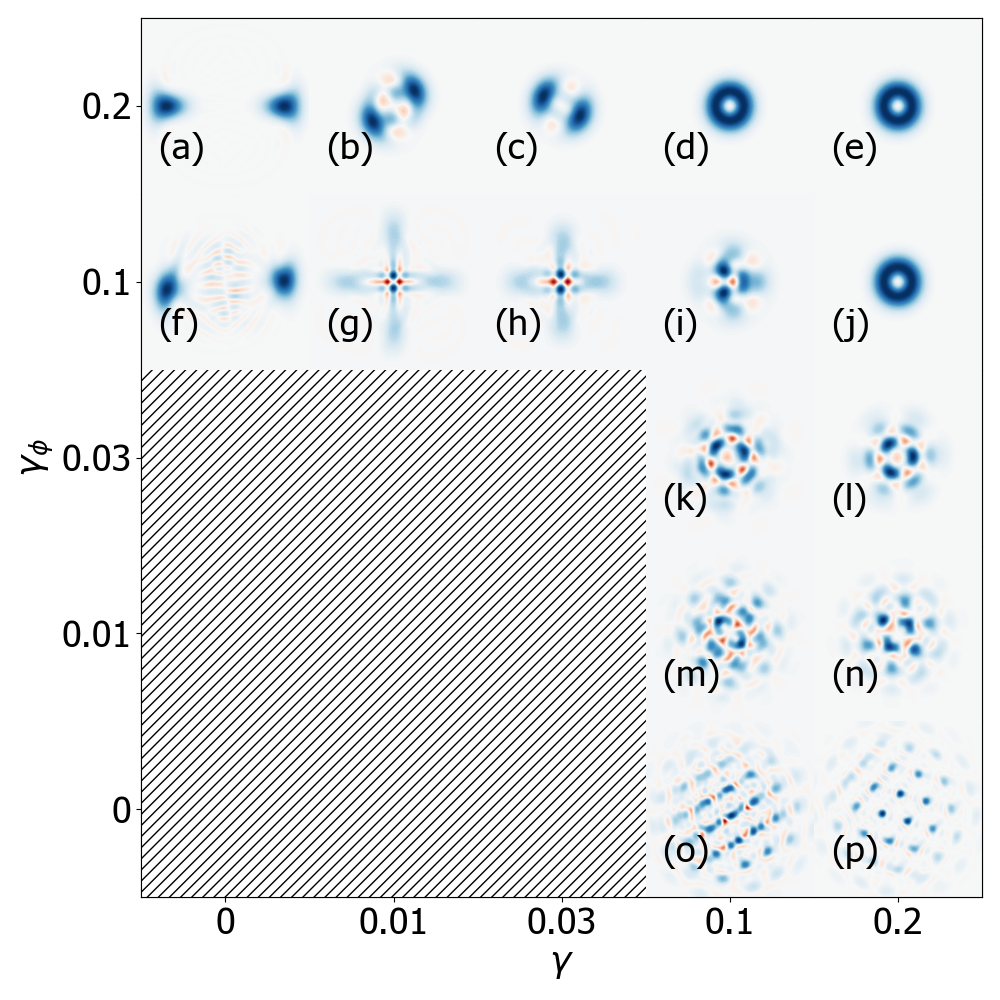}
  \caption{Wigner plots of the maximally mixed states for optimal codes. The codes are obtained using a biconvex optimization process for different rates of loss and dephasing under the energy constraint $\bar{n}\leq 9$. The plotted codes are consistently obtained from various randomly chosen initial codes. The shaded region represents a low-error range for which multiple local optima exist with entanglement fidelity approaching unity. The energy constraint is saturated when either the dephasing or loss rates are zero, namely for (a),(f),(o), and (p). The remaining codes have optimal mean energies, which we specify in Appendix \ref{app:optimization}.}
  \label{fig:phaseplot}
\end{figure}

\begin{figure*}[t]
  \includegraphics[width=\textwidth]{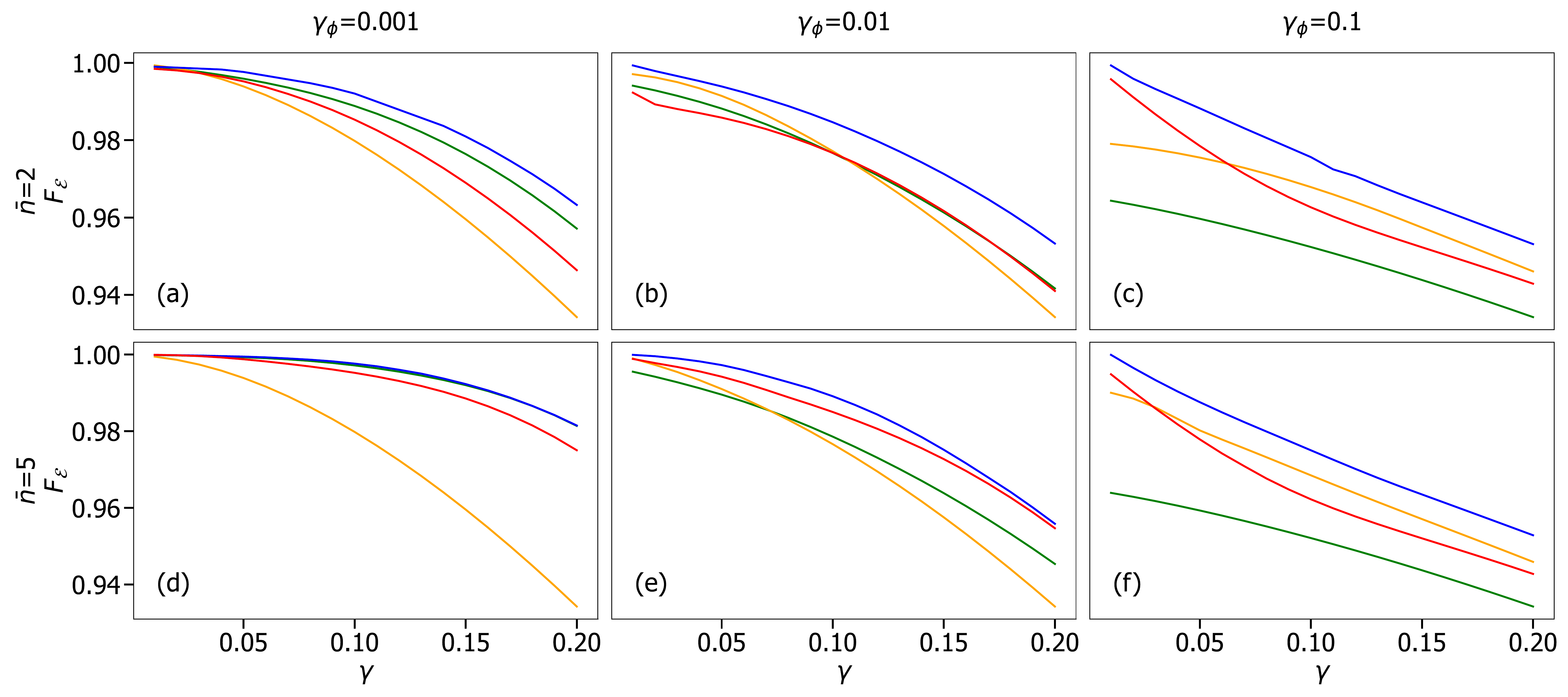}
  \caption{Entanglement fidelities $F_{\mathcal{E}}$ versus loss rate $\g$ for various dephasing rates $\gphi$ (columns) and energy constraints $\bar{n}$ (rows). The plots compare the performance of biconvex optimized qubit codes (blue) with optimal members of known qubit encoding families: four-legged cat qubit codes (orange), hexagonal GKP qubit codes (green) and numerical codes (red). In panel (d), the lines corresponding to GKP codes and the optimized codes nearly overlap.}
  \label{fig:entanglementfid}
\end{figure*}
The optimization results depicted in Fig. \ref{fig:phaseplot} demonstrate the interplay between loss and dephasing and indicate the ranges of $\g,\gphi$ in which each of the GKP, cat and numerical codes excel \cite{Joshi2021QuantumQED}. The optimal codes for pure-loss noise (Figs. 2(o) and 2(p)) closely resemble hexagonal GKP codes, as previously observed by Noh et al. \cite{Noh2020}. The optimal codes saturate the energy constraint, suggesting that GKP codes with more photons are better suited for dealing with loss. 

In the case of pure-dephasing noise (Figs. 2(a) and 2(f)), we observe that two-legged cat codes and squeezed two-legged cat codes \cite{Schlegel2022QuantumStates} provide optimal protection. Similarly to the GKP codes in the pure loss case, these codes saturate the energy constraint. This agrees with the known fact that the performance of two-legged cat codes improves as their mean energy increases \cite{Grimm2020,Berdou2022OneOscillator,Mirrahimi2014,Lescanne2020}. Squeezed two-legged cat codes have also been found to provide limited protection against photon loss in addition to dephasing \cite{Schlegel2022QuantumStates}.

Figs. 2(g) and 2(h) correspond to an encoding in which the code words are two-legged cats with opposite parities and orthogonal orientations in phase space. This encoding differs from the regular two-legged cat code, in which the orientation of the code words is identical. The separation of the code words in phase space offers protection against single-photon loss, besides the code's suppression of dephasing errors (a similar observation was made in Ref. \cite{Li2021Phase-engineeredCodes}). The modified two-legged cat code is also superior to the four-legged cat code \cite{Ofek2016} in that it also makes use of odd Fock states.

For all codes with nonzero dephasing and loss (i.e., $\g\geq 0.01,\gphi\geq 0.1$ or $\gphi\geq0.01,\g\geq 0.1$), we observe that the energy constraint  $\bar{n}\leq 9$ is not saturated (see Table \ref{MeanEnergies:1} in Appendix \ref{app:optimization}). Instead, these codes are characterized by varying optimal mean energies for different error rates. This situation is similar to the family of numerical codes \cite{Michael2016NewMode}, which consists of five different codes categorized by mean occupation number. However, unlike the numerical codes, our optimization process considers the error rates, resulting in different optimal codes for different $\g,\gphi$. Out of those, (i),(k),(l) and (n) appear to be similar to some of the numerical codes (see Fig. \ref{fig:numericcodes} in the Appendix). 

We also observe that certain codes have explicit rotational symmetry. For example, Figs. 2(k) and 2(l) are symmetric with respect to rotations by $2\pi/3$. However, these codes are not rotation codes \cite{Grimsmo2020}. Indeed, whereas the Fock-state distribution of the code words obeys $2N$-modularity, the remainders are not restricted to $0$ and $N$ as is the case for rotation codes (see Appendix \ref{rotationsymm}).

\subsection{\label{entfidel} Entanglement fidelity} 
To evaluate the performance of the numerically optimized qubit codes, we compare their entanglement fidelities to those of the other major code families. Specifically, we use the following codes: hexagonal GKP codes as representatives of GKP codes, four-legged cat codes as representatives of rotation codes and finally, the five numerical codes presented in Ref. \cite{Albert2018}.

Within each of the code families, we choose the particular instance of the family that maximizes the entanglement fidelity, i.e.,
\begin{equation}\label{explanation_fidelity_family}
        \underset{\mathcal{S}}{\text{argmax}}\{F_{\mathcal{R}^*\circ\lossdeph{\g}{\gphi}\circ\mathcal{S}}\},
\end{equation}
where $\mathcal{S}$ is a four-legged cat, hexagonal GKP or numerical code qubit encoding with $\mathrm{Tr}(\hat{n}\mathcal{S}(\h{\pi}))\leq\bar{n}$, and $\mathcal{R}^*=\mathrm{argmax}_\mathcal{R}\,F_{\mathcal{R}\circ\lossdeph{\g}{\gphi}\circ\mathcal{S}}$ is an optimal decoding channel.
The results of this comparison are shown in Fig. \ref{fig:entanglementfid}. In all cases, we find that the numerically optimized codes have the highest entanglement fidelity, as expected. For low dephasing rates ($\gphi=0.001$), hexagonal GKP codes offer the highest fidelity among the considered code families, while for high dephasing rates ($\gphi=0.1$) cat codes perform better. The numerical codes offer good results for intermediate dephasing $\gphi=0.01$ and sufficiently large $\nbar$.

\begin{figure*}[t]
  \includegraphics[width=\textwidth]{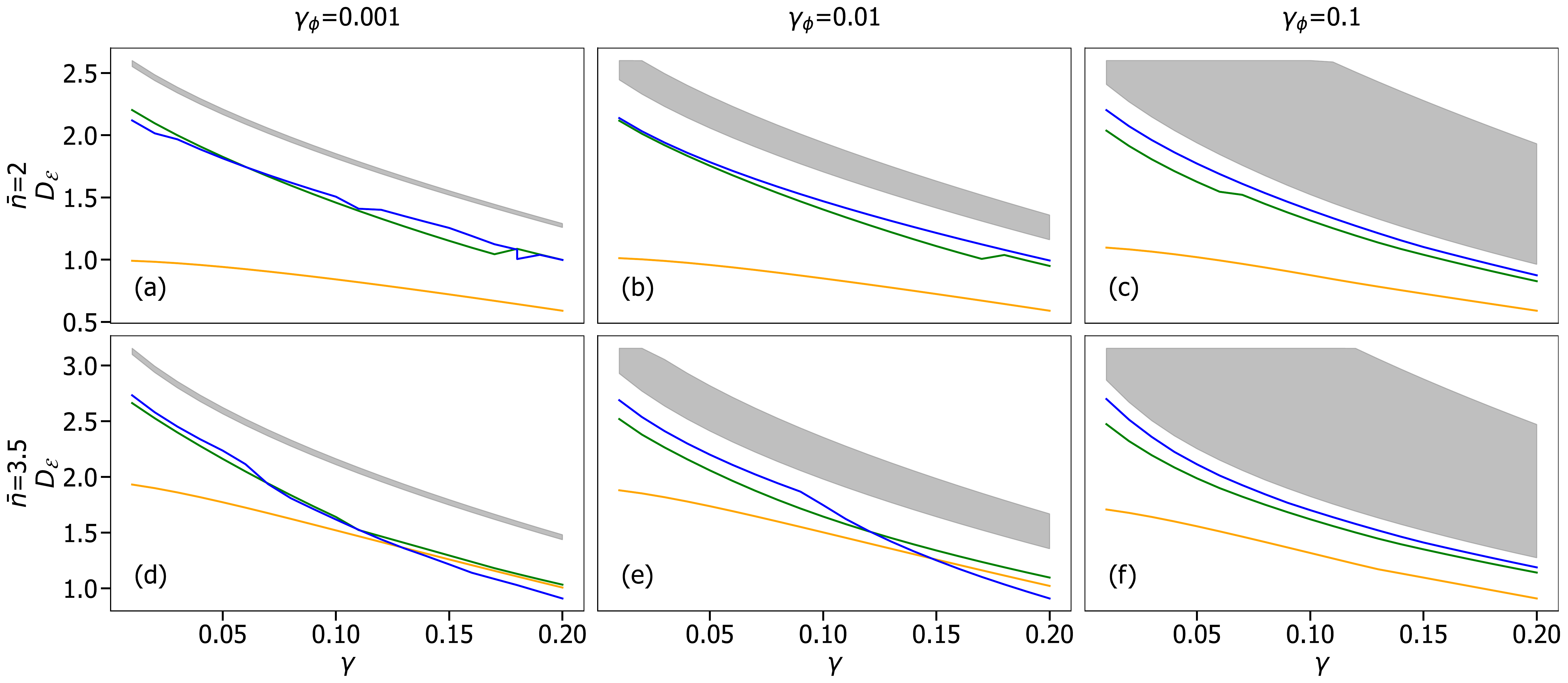}
  \caption{Comparison of the optimal hashing bounds $D_{\mathcal{E}}$ versus loss rate $\g$ for various dephasing rates $\gphi$ (columns) and energy constraints $\bar{n}$ (rows). The blue lines correspond to biconvex optimized qudit codes, the orange lines to the 2$d$-legged cat qudit codes, and the green lines to the hexagonal GKP qudit codes. The gray region corresponds to the capacity bounds presented in Eqs. \ref{eq:data_processing_bound}, \ref{eq:lower_bound}. Due to limitations in numerical convergence, this figure uses qudit dimension $d\leq 8$ and $\bar{n}=3.5$ as the higher energy constraint (instead of $\bar{n}=5$ as in the other figures).}
  \label{fig:hashing}
\end{figure*}

\subsection{\label{hashing} Hashing bound} 
The bosonic codes derived in the previous section can be used to reliably communicate information over the loss-dephasing channel. In this section, we study how close the communication rate of each code comes to the quantum capacity of the channel $Q_{\mathcal{N}}$.
While the achievable communication rate $Q_{\mathcal{E}}$ of an encoding-noise-decoding channel $\mathcal{E}$ is difficult to determine, it is bounded from below by the information-theoretic quantity known as the hashing bound $D_{\mathcal{E}}$, which we define below (see also corollary 21.2.1 and theorem 24.9.1 in  \cite{Wilde2013} and theorem 3.1 in \cite{Devetak2003DistillationStates}).

Let $\Cha{S}{M_A}{A},\Cha{R}{B}{M_B}$ (with $M_A$ and $M_B$ isomorphic) be an encoding-decoding pair for the channel $\Cha{N}{A}{B}$, and let $M'$ be a shared system isomorphic to $M_A,M_B$. If $\Newket{\Gamma}_{M_AM'}=\frac{1}{\sqrt{\dim M_A}}\sum_{i}\Newket{ii}$ is a maximally entangled state and $\hrho_{\mathcal{E}}=(\mathcal{E}\otimes\mathcal{I})(\Newketbra{\Gamma}{\Gamma})\in\mathcal{D}(M_BM')$, then the hashing bound $D_{\mathcal{E}}$ of the combined channel $\Cha{E}{M_A}{M_B}=\Cha{R}{B}{M_B}\circ\Cha{N}{A}{B}\circ\Cha{S}{M_A}{A}$ is defined as  
\begin{equation}
    D_{\mathcal{E}} = H(\NewTr{M'}{\hrho_{\mathcal{E}}})-H(\hrho_{\mathcal{E}}).
\end{equation}

From the definition, we see that $D_{\mathcal{E}}\leq \log_2\dim{\Hil{M'}}$. Therefore, to meaningfully compare the communication rates of the codes to the theoretical limit $Q_{\mathcal{N}}$, we generalize the qubit codes used earlier to qudit codes with dimension $d=\dim{\Hil{M'}}\geq 2^{Q_{\mathcal{N}}}$. 
Specifically, we use qudit codes obtained through biconvex optimization of the entanglement fidelity and known qudit code families, such as hexagonal GKP qudits (Eq. \ref{hexgkpqudit} in the Appendix) and 2$d$-legged cat qudits (Eq. \ref{2dcat}). 

In Fig. \ref{fig:hashing}, we plot the optimal hashing bounds
\begin{equation}\label{explanation_hashing_family}
\max_{\mathcal{S}}\{D_{\mathcal{R}^*\circ\lossdeph{\g}{\gphi}\circ\mathcal{S}}\},
\end{equation}
where $\mathcal{S}$ are the qudit encodings for the respective code families with $8\geq d\geq 2$ and mean energy $\NewTr{}{\hat{n}\mathcal{S}\brackets{\h{\pi}}}\leq\nbar$, and $\mathcal{R}^*=\mathrm{argmax}_\mathcal{R}\,F_{\mathcal{R}\circ\lossdeph{\g}{\gphi}\circ\mathcal{S}}$ are the optimal decodings.

The results indicate that the biconvex optimized codes and the GKP codes have very similar hashing bounds for all the considered parameter regimes. In addition, the hashing bounds are close to the previously derived lower bounds on the capacity of the loss-dephasing channel (Fig. 1). In some cases, we observe that the hashing bound of the hexagonal GKP code surpasses that of the biconvex optimized code. As previously observed in Ref. \cite{Albert2018}, optimizing the code for entanglement fidelity, as in our biconvex optimization scheme, does not necessarily imply an optimal hashing bound \cite{Bausch2018QuantumNetworks}. For example, introducing noise structure (e.g., biased noise) to $\mathcal{E}$ at the cost of slightly reduced entanglement fidelity after decoding the inner bosonic code may result in a more efficient outer code using multiple copies of $\mathcal{E}$. This procedure might lead to a better hashing bound than the one associated with the bosonic code achieving maximal entanglement fidelity.

Finally, the relatively low hashing bound of the cat qudit code is due to the fact that for $k=0,\ldots,d-1$ the code word $\Newket{k}_{\text{cat}}$ has support on Fock states $2k$ modulo $2d$ (see section \ref{rotationsymm} in the Appendix). Therefore, the mean energy of the code is greater than $d-1$, limiting the hashing bound to $\log_2 \nbar$.  
This implies that rotation codes are less compressed than other codes and require higher mean energy to provide the same capacity.

\section{\label{conclusion} Conclusion} 
We presented a study of the bosonic loss-dephasing channel from multiple perspectives. We showed that, unlike the pure-loss (for $\gamma\leq \frac{1}{2}$) and pure-dephasing channels, the loss-dephasing channel is not degradable, complicating the calculation of its capacity. To that end, we provided upper and lower bounds that prove to be tight for realistic values of $\g,\gphi$. Next, we used a biconvex optimization scheme with an energy constraint to find numerically optimized codes for various loss and dephasing rates. We observed that two-legged cat codes are well suited for dephasing errors, while hexagonal GKP codes handle loss errors well. These two edge cases saturate the energy constraint. In contrast, when both loss and dephasing are present, the energy constraint is not saturated, and codes resembling numerical codes emerge. The optimization procedure reveals a phase space of codes that vary non-smoothly from GKP codes to cat codes. Finally, we connected the two perspectives using the hashing bound and showed that the single-mode biconvex optimized codes give rise to a satisfactory lower bound on the capacity. This implies that the optimized codes can be used for quantum communication schemes over the loss-dephasing channel with a relatively high communication rate.
\\
\indent The remaining open questions include a study of whether or not the channel is anti-degradable in a nontrivial error range (not just for $\gamma\geq\frac{1}{2}$). We conjecture that the channel is not anti-degradable for $\gamma<\frac{1}{2}$ and suggest that this may be proven using similar methods to the ones used here. Furthermore, good analytical bounds on the capacity remain to be found. One may also attempt to prove or contradict that for any $\g,\gphi>0$, there exists a code with finite energy that maximizes the entanglement fidelity of $\lossdeph{\g}{\gphi}$ and, if so, estimate its energy. Finally, this work does not consider the implementation of encoding, error correction, and decoding procedures. However, our results may be a good starting point for finding codes that are both experimentally feasible and perform well under realistic loss-dephasing noise.
\begin{acknowledgments}

L.J. and Q.X. acknowledge support from the ARO (W911NF-18-1-0020, W911NF-18-1-0212), ARO MURI (W911NF-16-1-0349, W911NF-21-1-0325), AFOSR MURI (FA9550-19-1-0399, FA9550-21-1-0209), AFRL (FA8649-21-P-0781), DoE Q-NEXT, NSF (OMA-1936118, EEC-1941583, OMA-2137642), NTT Research, and the Packard Foundation (2020-71479). S.R. and P.L. acknowledge
financial support by the Israel Science Foundation and the European Research Council
starting investigator grant Q-CIRC 134847.

\end{acknowledgments}
\appendix
\begin{widetext}
\section{Quantum capacity of the pure-dephasing channel\label{app:dephasing}} The energy-constrained quantum capacity of the dephasing channel can be evaluated using a numerical convex optimization procedure. The optimization can be limited to diagonal states due to the covariance of this channel with respect to phase-space rotations. We observe that the photon number distribution of the optimal input state resembles the Poissonian distribution of coherent states in shape, although it may have a different variance  (see Fig. \hyperref[fig:optimizeddeph]{5}). Interestingly, for a large range of parameters, the coherent information of thermal states $\hat{\tau}(\nbar)$ yields an excellent approximation to the quantum capacity (see Fig. \hyperref[fig:optimizeddephvsthermal]{6}).

\begin{figure}[h]
\centering
\begin{minipage}{.44\textwidth}
\justifying
  \includegraphics[width=.95\linewidth]{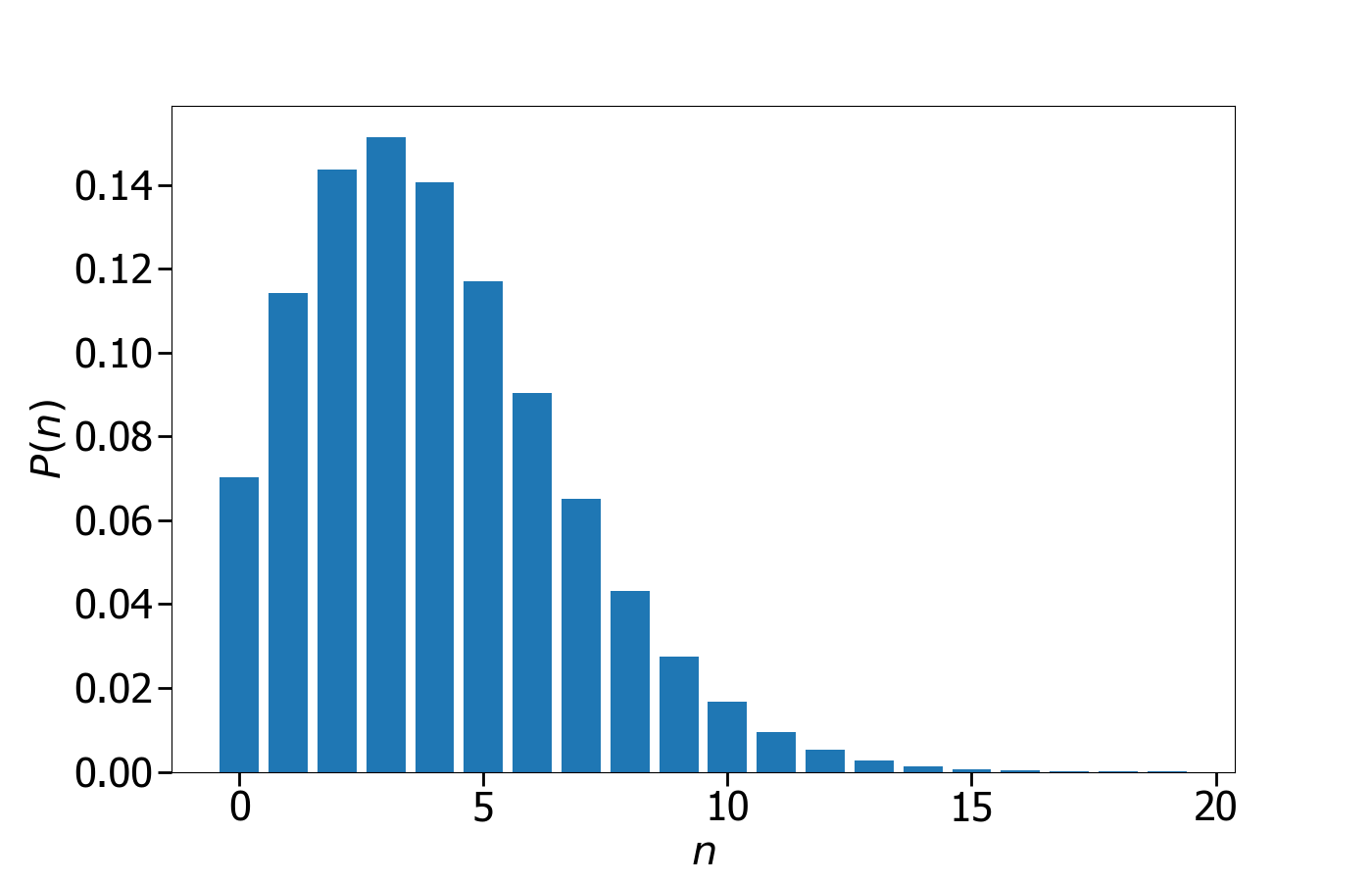} 
  \caption{Photon number distribution of an optimal \\diagonal state for pure-dephasing noise with $\gphi=0.1$ \\and energy constraint $\nbar=4$.}
  \label{fig:optimizeddeph}
\end{minipage}%
\begin{minipage}{.46\textwidth}
  \justifying
  \includegraphics[width=.9\linewidth]{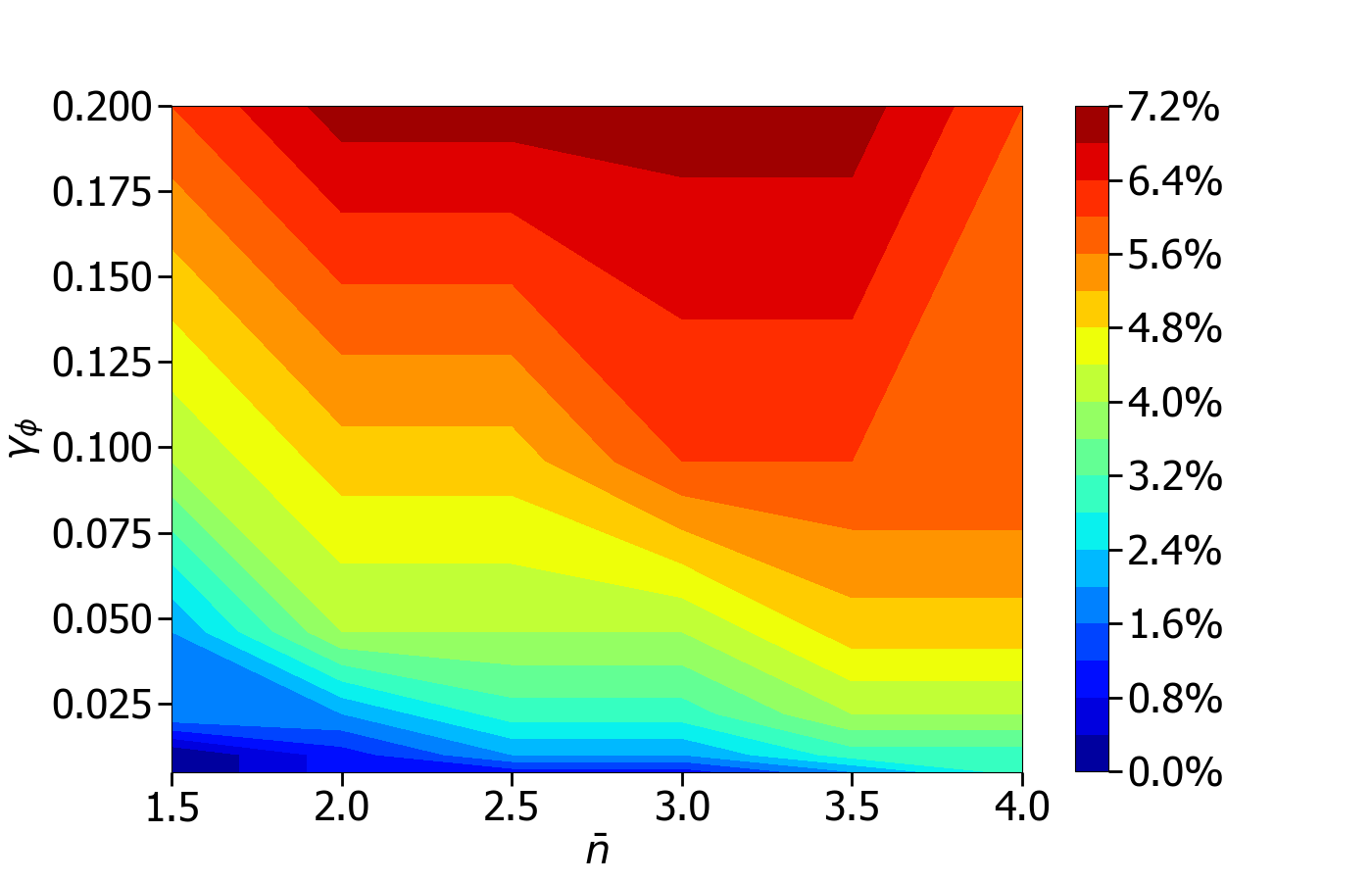} 
  \caption{Relative difference (in \%) in coherent information of thermal and optimal states as a function of dephasing rate $\gphi$ and energy constraint $\nbar$.}
  \label{fig:optimizeddephvsthermal}
\end{minipage}
\end{figure}

\section{\label{sec:nondegradabilityappendix} Non-degradability of the joint bosonic
loss and dephasing channel}

The aim of this appendix is to give a rigorous proof of the non-degradability of the loss-dephasing channel (Theorem \ref{nondegradabilitythm}) through a number of lemmas.

\begin{lemma}\label{lem:compchannel}
Let $U=e^{\sqrt{\gphi}\num{a}\brackets{\cre{b}_d-\ann{b}_d}}e^{\Newasin{\sqrt{\g}}\brackets{\cre{b}_l\h{a}-\cre{a}\ann{b}_l}}$ be the Stinespring dilation of a loss-dephasing channel $\lossdeph{\g}{\gphi}$. The channel $\lossdeph{\g}{\gphi}$ is then given by 
\begin{equation}
    \forall n,m\geq0,\,\lossdeph{\g}{\gphi}\brackets{\Newketbra{n}{m}}=e^{-\frac{1}{2}\gphi(n-m)^2}
    \sum_{\overset{k,l=0}{n-m=k-l}}^{n,m}\sqrt{\binom{n}{k}\binom{m}{l}(1-\g)^{k+l}\g^{n+m-k-l}}\Newketbra{k}{l}
\end{equation}
and the 
complementary channel $\lossdephcomp{\g}{\gphi}$ is given by 
\begin{equation}
    \forall n,m\geq0,\,\lossdephcomp{\g}{\gphi}\brackets{\Newketbra{n}{m}}=\sum_{k=0}^{\min(n,m)}\sqrt{\binom{n}{k}\binom{m}{k}\g^{n+m-2k}(1-\g)^{2k}}\Newketbra{n-k,\sqrt{\gphi}k}{m-k,\sqrt{\gphi}k}.
\end{equation}
\end{lemma}
\begin{proof}
We start by calculating the state $U\Newket{n00}_{XE_lE_d}$ for any $n\geq0$. We can combine the two identities
\begin{equation}
    e^{\Newasin{\sqrt{\g}}\brackets{\cre{b}\h{a}-\cre{a}\ann{b}}}\Newket{n0}=\sum_{k=0}^n\sqrt{\binom{n}{k}(1-\g)^{k}\g^{n-k}}\Newket{k,n-k}
\end{equation}
\begin{equation}
e^{\sqrt{\gphi}\num{a}\brackets{\cre{b}-\ann{b}}}\Newket{n0}=\Newket{n,\sqrt{\gphi}n}
\end{equation}

to determine that $\forall n\geq 0$,
\begin{equation}
    U\Newket{n00}_{XE_lE_d}=\sum_{k=0}^n\sqrt{\binom{n}{k}(1-\g)^k\g^{n-k}}\Newket{k,n-k,\sqrt{\gphi}k}_{XE_lE_d}.
\end{equation}

We can now calculate the isometry on an arbitrary operator basis element $\Newketbra{n}{m}$:
\begin{equation}
    U\Newketbra{n00}{m00}U^{\dag}=\sum_{k,l=0}^{n,m}\sqrt{\binom{n}{k}\binom{m}{l}(1-\g)^{k+l}\g^{n+m-k-l}}\Newketbra{k,n-k,\sqrt{\gphi}k}{l,m-l,\sqrt{\gphi}l}.
\end{equation}

Finally, tracing over the system $X$ forces $k=l$, which yields the desired result for the complementary channel:
\begin{equation}
\begin{split}
    \lossdephcomp{\g}{\gphi}\brackets{\Newketbra{n}{m}}&=\NewTr{X}{U\Newketbra{n00}{m00}U^{\dag}}\\&=\sum_{k=0}^{\min(n,m)}\sqrt{\binom{n}{k}\binom{m}{k}\g^{n+m-2k}(1-\g)^{2k}}\Newketbra{n-k,\sqrt{\gphi}k}{m-k,\sqrt{\gphi}k}.
\end{split}
\end{equation}

Similarly, tracing over the systems $E_d,E_l$ forces $n-m=k-l$. Using  the overlap between the coherent states $\Newbraket{\sqrt{\gphi}k}{\sqrt{\gphi}l}=e^{-\frac{1}{2}\gphi(l-k)^2}$, we obtain the expression for $\lossdeph{\g}{\gphi}$.
\end{proof}

Next, if we assume by contradiction that the loss-dephasing channel is degradable, then the degrading channel must assume the form presented in the following lemma.

\begin{lemma}\label{lem:degrading}
If $\g<1$ and $\mathcal{D}$ is a degrading super-operator s.t. $\mathcal{D}\circ\lossdeph{\g}{\gphi}=\lossdephcomp{\g}{\gphi}$ then $\mathcal{D}$ must be the super-operator defined by 
\begin{equation}\label{eq:degradation_super}
\begin{split}
    &\forall n,m\geq0,\,\mathcal{D}\brackets{\Newketbra{n}{m}}=\\&e^{\frac{1}{2}\gphi(n-m)^2}\sum_{t=0}^{\min(n,m)}\sum_{l=0}^{t}(-1)^{t-l}\sqrt{\binom{n}{n-t,l}\binom{m}{m-t,l}\brackets{\frac{\g}{1-\g}}^{n+m-2l}}\Newketbra{n-t,\sqrt{\gphi}l}{m-t,\sqrt{\gphi}l}.
\end{split}
\end{equation}
If $\g=1$, no such super-operator exists.
\end{lemma}
\begin{proof}
We start by considering maximum photon loss, i.e., $\g=1$. In this case, the channel $\lossdeph{\g}{\gphi}$ maps all states to the vacuum state. Since the complementary channel $\lossdephcomp{\g}{\gphi}=\mathcal{I}_{X\to E_l}\otimes \Newketbra{0}{0}_{E_d}$ is not a constant channel, the degrading map $\mathcal{D}$ does not exist.

Now, assume that $\g<1$. Then, for all integer $\Delta,n\geq 0$,
\begin{equation}
    \lossdephcomp{\g}{\gphi}\brackets{\Newketbra{n}{n+\Delta}} = \mathcal{D}\circ\lossdeph{\g}{\gphi}\brackets{\Newketbra{n}{n+\Delta}}.
\end{equation}

Using the previous lemma, we obtain
\begin{equation*}
\begin{split}
    \lossdephcomp{\g}{\gphi}\brackets{\Newketbra{n}{n+\Delta}}  = & \sum_{k=0}^{n}\sqrt{\binom{n}{k}\binom{n+\Delta}{k}\g^{2n+\Delta-2k}(1-\g)^{2k}}\Newketbra{n-k,\sqrt{\gphi}k}{n+\Delta-k,\sqrt{\gphi}k} \\ = & e^{-\frac{1}{2}\gphi\Delta^2}
    \sum_{k=0}^{n}\sqrt{\binom{n}{k}\binom{n+\Delta}{k+\Delta}(1-\g)^{2k+\Delta}\g^{2n-2k}}\mathcal{D}\brackets{\Newketbra{k}{k+\Delta}} \\ = &
    \mathcal{D}\circ\lossdeph{\g}{\gphi}\brackets{\Newketbra{n}{n+\Delta}}.
\end{split}
\end{equation*}
Using the abbreviated notation
\begin{equation*}
    \begin{split}
        X_{k,\Delta} &= e^{-\frac{1}{2}\gphi \Delta^2}\frac{(1-\g)^{k+\frac{\Delta}{2}}}{\sqrt{k!(k+\Delta)!}\g^k}\mathcal{D}\brackets{\Newketbra{k}{k+\Delta}}\\
        Y_{n,\Delta} &= \sum_{k=0}^n\sqrt{\frac{1}{k!^2(n-k)!(n+\Delta-k)!}\g^{\Delta-2k}(1-\g)^{2k}}\Newketbra{n-k,\sqrt{\gphi}k}{n+\Delta-k,\sqrt{\gphi}k},
    \end{split}
\end{equation*}
the previous equation simplifies to $\sum_{k=0}^n\frac{X_{k,\Delta}}{(n-k)!}=Y_{n,\Delta}$. This allows one to invert the expression and obtain $X_{n,\Delta}=\sum_{k=0}^n\frac{(-1)^{n-k}Y_{k,\Delta}}{(n-k)!}$ by induction on $n$. Using the definition of $X_{k,\Delta}$ we can rewrite $\mathcal{D}(\Newketbra{n}{n+\Delta})$ as 
\begin{equation}
    \begin{split}
        &\mathcal{D}\brackets{\Newketbra{n}{n+\Delta}} =  e^{\frac{1}{2}\gphi\Delta^2}\brackets{\frac{(1-\g)^{n+\Delta/2}}{\sqrt{n!(n+\Delta)!}\g^n}}^{-1}\times\\&\sum_{k=0}^n\sum_{l=0}^k\frac{(-1)^{n-k}}{(n-k)!}\sqrt{\frac{1}{l!^2(k-l)!(k+\Delta-l)!}\g^{\Delta-2l}(1-\g)^{2l}}\Newketbra{k-l,\sqrt{\gphi}l}{k+\Delta-l,\sqrt{\gphi}l}
        =\\ & e^{\frac{1}{2}\gphi\Delta^2}\sum_{k=0}^n\sum_{l=0}^k(-1)^{n-k}\sqrt{\binom{n}{k-l,l}\binom{n+\Delta}{k-l+\Delta,l}\brackets{\frac{\g}{1-\g}}^{2n-2l+\Delta}}\Newketbra{k-l,\sqrt{\gphi}l}{k+\Delta-l,\sqrt{\gphi}l}=\\&\text{ (take }t=n-k+l\text{)}=\\ & e^{\frac{1}{2}\gphi\Delta^2}\sum_{t=0}^n\sum_{l=0}^t(-1)^{t-l}\sqrt{\binom{n}{n-t,l}\binom{n+\Delta}{n+\Delta-t,l}\brackets{\frac{\g}{1-\g}}^{2n-2l+\Delta}}\Newketbra{n-t,\sqrt{\gphi}l}{n+\Delta-t,\sqrt{\gphi}l}.
    \end{split}
\end{equation}
Taking the adjoint of both sides of the equation gives us an expression for $\mathcal{D}\brackets{\Newketbra{n+\Delta}{n}}$, so that Eq. \ref{eq:degradation_super} holds, as required.
\end{proof}

To prove that the channel is not degradable, we will evaluate the degradation super-operator on coherent states and show that it can map them to non-positive operators. We start with the following lemma:
\begin{lemma}
For $\g<1$ and $\alpha\in\mathbb{C}$, the map $\mathcal{D}$ satisfies 
\begin{equation}
\mathcal{D}\brackets{\Newketbra{\alpha}{\alpha}}=\mathcal{N}_{-\gphi}\brackets{\Newketbra{\sqrt{\frac{\g}{1-\g}}\alpha}{\sqrt{\frac{\g}{1-\g}}\alpha}}\otimes\h{\sigma},    
\end{equation}
 where $\h{\sigma}=\sumzeroinfty{l}e^{-|\alpha|^2}\frac{|\alpha|^{2l}}{l!}\Newketbra{\sqrt{\gphi}l}{\sqrt{\gphi}l}$ is a density matrix and $\mathcal{N}_{-\gphi}=\exp\brackets{-\gphi\brackets{\n\bullet\n-\frac{1}{2}\{\n^2,\bullet\}}}$ is a dephasing super-operator with negative dephasing rate, i.e., it is not a channel.
\end{lemma}
\begin{proof}
Plugging in the expression for $\mathcal{D}$ from the previous lemma and using $\Newketbra{\alpha}{\alpha}=e^{-|\alpha|^2}\sum_{n,m=0}^{\infty}\frac{\alpha^n\bar{\alpha}^m}{\sqrt{n!m!}}\Newketbra{n}{m}$, we obtain 
\begin{equation*}
    \begin{split}
        \mathcal{D}\brackets{\Newketbra{\alpha}{\alpha}} = & \sumzeroinfty{n}\sumzeroinfty{m}e^{\frac{1}{2}\gphi(n-m)^2}\sum_{t=0}^{\min(n,m)}\sum_{l=0}^{t}(-1)^{t-l}\Bigg(\sqrt{\binom{n}{n-t,l}\binom{m}{m-t,l}\brackets{\frac{\g}{1-\g}}^{n+m-2l}}\times\\\times&e^{-|\alpha|^2}\frac{\alpha^n\bar{\alpha}^m}{\sqrt{n!m!}}\Newketbra{n-t,\sqrt{\gphi}l}{m-t,\sqrt{\gphi}l}\Bigg)\\ = &
        \sumzeroinfty{n}\sumzeroinfty{m}e^{\frac{1}{2}\gphi(n-m)^2-|\alpha|^2}\sum_{t=0}^{\min(n,m)}\Bigg(\frac{|\alpha|^{2t}}{t!}\frac{\brackets{\sqrt{\frac{\g}{1-\g}}\alpha}^{n-t}}{\sqrt{(n-t)!}}\frac{\brackets{\sqrt{\frac{\g}{1-\g}}\bar{\alpha}}^{m-t}}{\sqrt{(m-t)!}}\Newketbra{n-t}{m-t}\otimes\\\otimes&\sum_{l=0}^{t}\brackets{\frac{\g}{\g-1}}^{t-l}\binom{t}{l}\Newketbra{\sqrt{\gphi}l}{\sqrt{\gphi}l}\Bigg).
    \end{split}
\end{equation*}
Taking $r=n-t,s=m-t,q=t-l$ we obtain,
\begin{equation*}
    \begin{split}
        \mathcal{D}\brackets{\Newketbra{\alpha}{\alpha}} =& \squarebrackets{\sumzeroinfty{q}\frac{|\alpha|^{2q}}{q!}\brackets{\frac{\g}{\g-1}}^{q}}\times\\\times&\squarebrackets{\sum_{r,s=0}^{\infty}e^{\frac{1}{2}\gphi(r-s)^2}\frac{\brackets{\sqrt{\frac{\g}{1-\g}}\alpha}^r}{\sqrt{r!}}\frac{\brackets{\sqrt{\frac{\g}{1-\g}}\bar{\alpha}}^s}{\sqrt{s!}}\Newketbra{r}{s}}\otimes\squarebrackets{\sumzeroinfty{l}e^{-|\alpha|^2}\frac{|\alpha|^{2l}}{l!}\Newketbra{\sqrt{\gphi}l}{\sqrt{\gphi}l}}.
    \end{split}
\end{equation*}
Finally, note that $\sumzeroinfty{q}\frac{|\alpha|^{2q}}{q!}\brackets{\frac{\g}{\g-1}}^q=\exp\brackets{-\frac{\g}{1-\g}|\alpha|^2}$, yielding the required result.
\end{proof}
The last lemma, which concludes the proof, shows that $\alpha$ can be chosen s.t. $\mathcal{D}\brackets{\Newketbra{\alpha}{\alpha}}$ is not a state. Hence $\mathcal{D}$ is not a quantum channel.
\begin{lemma}\label{lem:notchannel}
If $0<\g<1$ and $\gphi>0$ then $\mathcal{D}$ is not a quantum channel.
\end{lemma}
\begin{proof}
It is sufficient to show that $\mathcal{D}$ is not a positive map. Let $n\in\mathbb{N},\alpha\in\mathbb{C}$ s.t. $\brackets{\sqrt{\frac{\g}{1-\g}}\alpha}^n=-1$, then for $\Newket{\phi}=\frac{1}{\sqrt{2}}\brackets{\Newket{00}+\Newket{n0}}$ we have 
\begin{equation}
\begin{split}
    \Newmel{\phi}{\mathcal{D}\brackets{\Newketbra{\alpha}{\alpha}}}{\phi} =& \frac{1}{2}e^{-\frac{\g}{1-\g}|\alpha|^2}\brackets{1+\frac{\lvert\sqrt{\frac{\g}{1-\g}}\alpha\rvert^{2n}}{n!}+2e^{\frac{1}{2}\gphi n^2}\mathrm{Re}\brackets{\frac{\brackets{\sqrt{\frac{\g}{1-\g}}\alpha}^n}{\sqrt{n!}}}}\Newmel{0}{\h{\sigma}}{0}\\=&\frac{1}{2}e^{-\frac{\g}{1-\g}|\alpha|^2}\brackets{1+\frac{1}{n!}-2e^{\frac{1}{2}\gphi n^2}\frac{1}{\sqrt{n!}}}\Newmel{0}{\h{\sigma}}{0}\underset{n\to{\infty}}{\longrightarrow}-\infty,
\end{split}
\end{equation}
since $\Newmel{0}{\h{\sigma}}{0}>0$. This shows that $\mathcal{D}\brackets{\Newketbra{\alpha}{\alpha}}\ngeq0$. Therefore, $\mathcal{D}$ is not a positive map and in particular, not a quantum channel.
\end{proof}
\pagebreak
\end{widetext}
\section{Biconvex optimization of QEC codes for the loss-dephasing channel\label{app:optimization}}
We use biconvex optimization to find a local maximum in a complicated landscape defined by the entanglement fidelity. For the process to converge, it must run for many iterations. Moreover, the point to which it converges depends on the starting point. 

For Fig.  \ref{fig:phaseplot}, we ran the optimization process for around 3000 iterations per plot. This process was repeated ten times with different randomized starting codes in each repetition. We then selected the optimal results for each pair of $\g,\gphi$, while ensuring that the results from other repetitions yield similar codes. This consistency criterion indicates that the local optimum might also be a global optimum. This procedure works well when either $\g$ or $\gphi$ are large. However, when both parameters are small, the landscape becomes shallow.  This makes it hard to converge to a single global maximum, leading to different optimization results for different starting codes. For this reason, we chose not to include the cases where both  $\gphi$ and $\g$ are below $0.1$. 

As presented in the main text, the optimal codes do not always saturate the energy constraint (see Table \ref{MeanEnergies:1}).
\begin{table}
\begin{tabular}{ |p{1.4cm}||p{1.4cm}|p{1.4cm}|p{1.4cm}|  }
 \hline
 \multicolumn{4}{|c|}{Mean energies} \\
 \hline
 Index& $\g$ &$\gphi$ &$\bar{n}$\\
 \hline
a&0.0&0.2&9.00\\
b&0.01&0.2&1.70\\
c&0.03&0.2&1.13\\
d&0.1&0.2&0.50\\
e&0.2&0.2&0.50\\
f&0.0&0.1&8.81\\
g&0.01&0.1&5.33\\
h&0.03&0.1&3.52\\
i&0.1&0.1&1.38\\
j&0.2&0.1&0.50\\
k&0.1&0.03&3.17\\
l&0.2&0.03&2.30\\
m&0.1&0.01&4.28\\
n&0.2&0.01&3.34\\
o&0.1&0.0&8.38\\
p&0.2&0.0&9.00\\
 \hline
\end{tabular}
\caption{Mean energies of the optimized codes from Fig. \ref{fig:phaseplot} with energy constraint $\bar{n}\leq 9$. The energy constraint is saturated or nearly saturated when either the dephasing or loss rates are zero, namely for (a),(f),(o), and (p). }
\label{MeanEnergies:1}
\end{table}
In some of the cases where either $\g=0$ or $\gphi=0$, the mean energy $\bar{n}$ comes close to the constraint but does not saturate it. This is most likely due to numerical limitations of the optimization and the fact that we used a limited number of energy levels ($\dim{\mathcal{H}_X}=22$).
\subsection{\label{app:regularizationcompbound}Regularization process for comparison of numerically optimal codes}
The loss-dephasing channel is invariant under phase-space rotations, which causes optimization results with different starting points to be similar up to rotations. To tackle this, we regularize the codes by concatenating them with $e^{i\theta\hat{n}}\bullet e^{-i\theta\hat{n}}$ and imposing a condition that a particular off-diagonal element in the density matrix of the maximally mixed state of the resulting code is  positive. Similarly, the pure-loss channel is invariant under rotations, and the same regularization is applied there. The pure-dephasing channel, however, is also invariant under diagonal unitaries. Therefore, in that case we regularize the code by concatenating it with $e^{i\hat{D}}\bullet e^{-i\hat{D}}$, where $\hat{D}$ is a diagonal matrix with real entries. These entries are chosen such that a single off-diagonal element in each row of the density matrix of the maximally mixed state becomes positive. This results in codes with plots such as (a) and (f) in Fig. \ref{fig:phaseplot}.

\subsection{\label{app:starcode} Optimization results for the pure-loss channel with low energy constraints.} 
Hexagonal GKP codes emerge as optimal codes for the pure-loss channel if the energy constraint is chosen to be sufficiently high. However, if the energy constraint is low (e.g., $\bar{n}=6$ or $\bar{n}=7$), and if the loss rate is sufficiently high (e.g., $\g=0.2$), the optimization result is a code with five-fold rotation symmetry (see Fig. \ref{fig:starcode}). 
A potential insight as to why this might be the case is that, like the GKP code, we can use a tiling to protect against random shifts. However, due to the energy constraint, the tiling is not shift-invariant and distorts as we move away from the origin. A hyperbolic plane may be a good model for this phenomenon, and this plane can be tiled by pentagrams, as we observe here.
\setcounter{figure}{6}
 \begin{figure}[h!]
  \includegraphics[width=0.45\textwidth]{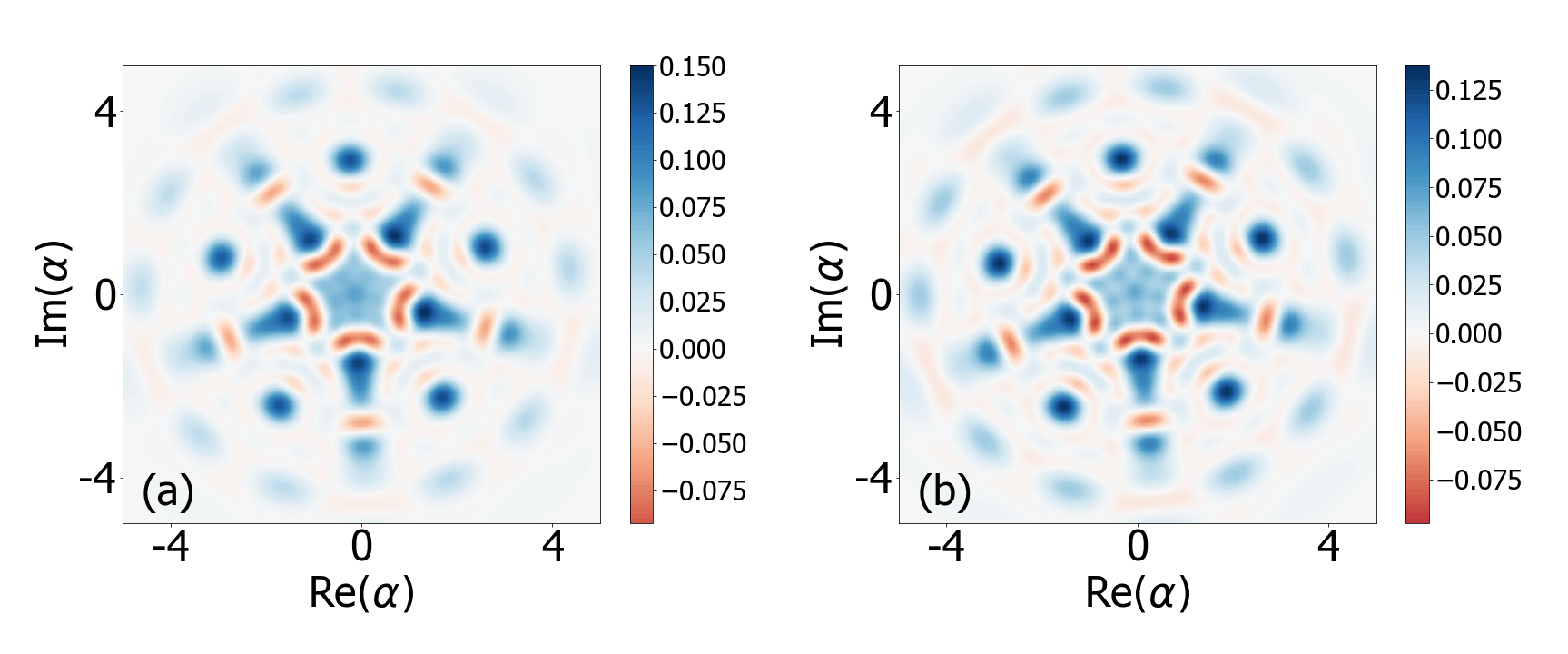}
  \caption{Wigner plots of codes with five-fold symmetry that are numerically optimized to handle pure loss. The codes are not translationally invariant and, therefore, do not qualify as GKP codes. The energy constraints are $\bar{n}=6$  (left plot) and $\bar{n}=7$ (right plot).}
  \label{fig:starcode}
\end{figure}

\section{Overview of bosonic error correction codes\label{app:knowncodes}}
\subsection{\label{gkpcode} GKP codes}
GKP codes \cite{Gottesman2001} are lattice codes that are particularly well suited for handling phase-space displacements and bosonic loss noise \cite{Albert2018}. GKP codes can be either single-mode (as used in this work) or defined using $N$ modes. If they are $2N$-dimensional, the code words can be seen as superpositions of infinitely many $N$-mode coherent states centered around points of a non-degenerate $2N$-dimensional lattice in phase space. As such, they have infinite energy. A common way to modify the code to have finite energy, is to multiply the coherent states by weights from a Gaussian envelope, with narrower envelopes corresponding to lower mean energy of the code. 

For example, if $V\in M_{2N}(\mathbb{C})$ is a matrix with columns $v_1,\ldots,v_{2N}$ that span a lattice $L$ in a $2N$-dimensional phase space with coordinates ($\h{q}_1,\ldots,\h{q}_N,\ldots\h{p}_1,\ldots,\h{p}_N$), we can define the code words of a finite-energy GKP qudit based on $V$ as \cite{Albert2018}
\begin{widetext}
\begin{equation}
    \begin{split}
        \Newket{\mu_{V,d}^{\Delta}}\propto & \sum_{\bar{n}=(n_1,\ldots,n_{2N})\in\mathbb{Z}^{2N}}e^{-\Delta^2 \Vert(dn_1+\mu)v_1+\sum_{i=2}^{2N}n_iv_i\Vert^2}\times \hat{D}_{(dn_1+\mu)v_1}\hat{D}_{n_2v_2}\ldots \hat{D}_{n_{2N}v_{2N}}\Newket{0\ldots 0},\,\mu=0,\ldots d-1,
    \end{split}
\end{equation}
\end{widetext}
where $\hat{D}_{(q_1,\ldots,q_n,p_1,\ldots p_n)}$ are $N$-mode displacements.

In particular, for the hexagonal GKP qudits $\mathcal{G}_{d,\Delta}$ ($\Newket{\mu_{\mathcal{G}_{d,\Delta}}}=\Newket{\mu_{V,d}^{\Delta}}$) referred to in the main text, the lattice is given by
\begin{equation}\label{hexgkpqudit}
    N=1,\, v_1=\sqrt{\frac{\pi}{d}}\sqrt{\frac{2}{\sqrt{3}}}\left(\frac{\sqrt{3}-i}{2}\right),\,v_2=\sqrt{\frac{\pi}{d}}\sqrt{\frac{2}{\sqrt{3}}}i.
\end{equation}

GKP codes can be seen as stabilizer codes over $2N$ commuting displacements $\hat{D}_{dv_1},\hat{D}_{v_2},\ldots,\hat{D}_{v_n}$. Therefore, not any lattice can be used to define a valid GKP code. In the single-mode case, for example, this requirement translates to $|\mathrm{det}(V)|=\frac{\pi}{2}$, so that a GKP code with square lattice will have $v_1=\sqrt{\frac{\pi}{2}}\begin{bmatrix}1\\0\end{bmatrix}$ and $v_2=\sqrt{\frac{\pi}{2}}\begin{bmatrix}0\\1\end{bmatrix}$. The condition on $V$ for $N>1$ is more involved and is discussed, for example, in section 2.4.3 of Ref. \cite{KyungjooNoh2020}.

\begin{figure}[h!]
  \includegraphics[width=0.4\textwidth]{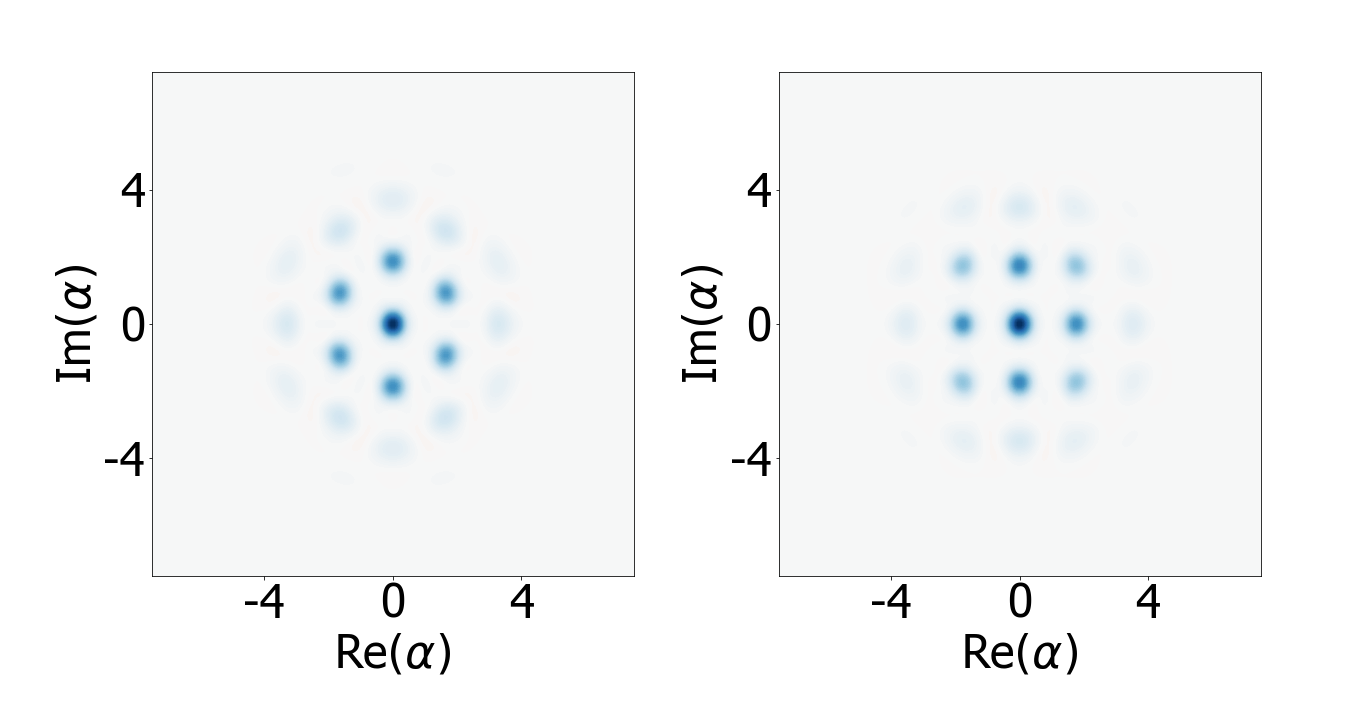}
  \caption{Wigner plots of the maximally mixed states for hexagonal (left panel) and square (right panel) lattice single-mode GKP codes.}
  \label{fig:gkpcode}
\end{figure}

The performance of a GKP code depends on the lattice that defines it, with lattices that have a higher ratio of sphere packing (ratio between the volume of the largest $2N$-sphere that can inhabit the Voronoi cell and the volume of the cell itself) performing better. This means, for example, that a single-mode GKP code defined by a hexagonal lattice will perform better than a code defined by a square lattice (see section 4 in Ref. \cite{Noh2019}). Both codes are presented in Fig. \ref{fig:gkpcode}.

Ref. \cite{Noh2019} also shows that GKP codes perform well for bosonic loss in two crucial ways. The first is that infinite-energy multi-mode GKP codes achieve a quantum communication rate that is offset by up to $\log_2 e$ from the capacity of the channel in the limit where the number of modes goes to infinity. This feature holds for any loss rate $\gamma$ between $0$ and $\frac{1}{2}$. The second advantage of GKP codes is that the biconvex optimization process described earlier usually converges to a state resembling a hexagonal GKP state with finite energy if only loss is present. However, as discussed in appendix \ref{app:optimization}, for certain energy constraints it can also converge consistently to other codes. For example, a code with five-fold symmetry can emerge, probably due to a requirement to tile a bounded section of phase space.

\subsection{\label{rotationsymm} Rotation codes} 
\begin{figure}[h]
\centering
  \includegraphics[width=0.45\textwidth]{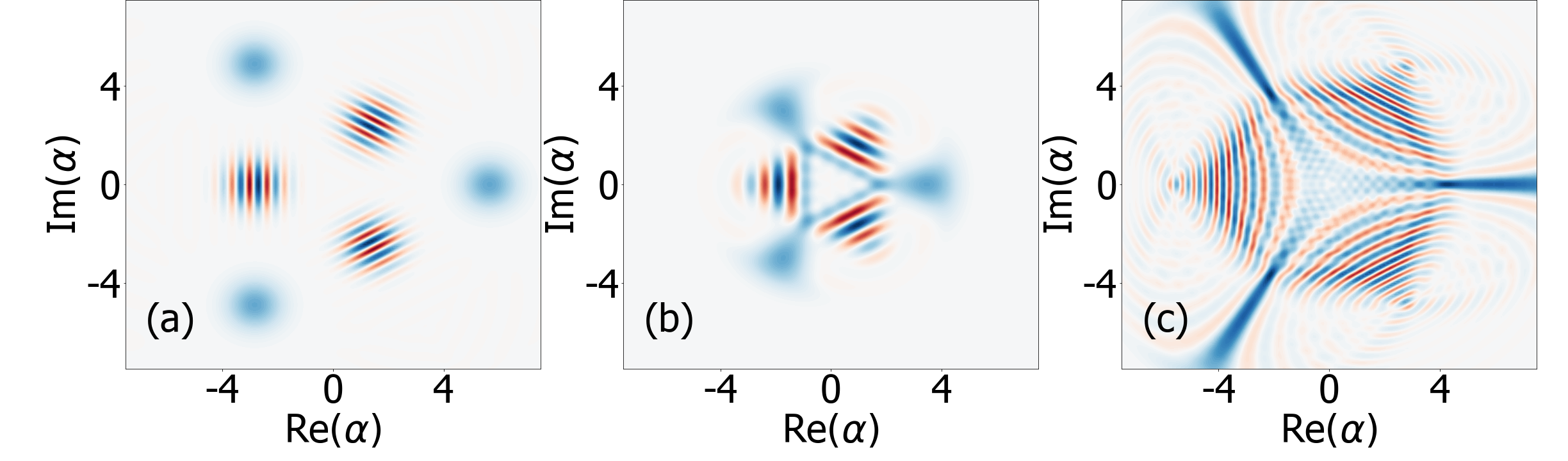}
  \caption{Examples of rotation codes with symmetry $N=3$. We show the Wigner distributions of the maximally mixed state for the cat code (left plot), the binomial code (middle plot), and the Pegg-Barnett code (right plot). }
  \label{fig:rotationcodesexample}
\end{figure}

A single-mode bosonic qudit encoding with code words $\Newket{k}_\mathcal{C},k=0,\ldots,d-1$ is a called a rotation code \cite{Grimsmo2020} with symmetry $N$ if the code words are invariant under the rotation $\hat{R}_{N}=e^{\frac{2i\pi}{N}\n}$ and if the rotation $\hat{R}_{dN}=e^{\frac{2i\pi}{dN}\n}$ acts as a logical $\h{Z}$ on the qudit: 
\begin{equation}
\begin{split}
    \forall k=0,\ldots d-1,& \hat{R}_{N}\Newket{k}_\mathcal{C}=\Newket{k}_\mathcal{C},\\&\hat{R}_{dN}\Newket{k}_\mathcal{C}=e^{\frac{2i\pi k}{N}}\Newket{k}_\mathcal{C}.
\end{split}
\end{equation}
Cat codes and binomial codes are examples of rotation codes (Fig. \ref{fig:rotationcodesexample}). Ref. \cite{Grimsmo2020} provides a scheme for performing encoding, decoding and gates on such codes based on controlled rotation gates and phase measurements.

The logical code words of a $2d$-legged cat qudit, as referred to in the main text, are defined by 
\begin{equation}\label{2dcat}
\Newket{k}_{\mathcal{C}_{d,\alpha}} \propto \sum_{l=0}^{2d-1}e^{-\frac{2\pi lki}{d}}\hat{R}_{2d}^{l}\Newket{\alpha}
\end{equation}
where $\Newket{\alpha}$ is a coherent state with $\alpha\neq0$.

Since the code word $\Newket{k}_\mathcal{C}$ of a rotation code is supported by Fock states that are $kN$ modulo $dN$, rotation codes provide protection (or at least detection) against up to $N$ photon loss events. If the code is also a number-phase code, that is, the phases of the code words in the conjugate basis are well localized (modulo a rotation), then the phases of the code words are also well separated. This property protects the code against dephasing errors. Number-phase codes are similar to GKP codes in that, whereas the former use photon number-phase duality to define the code, the latter rely on position-momentum duality.

 In our numerical optimizations, we found that for pure dephasing ($\gamma=0$), we converge to $N=1$ rotation codes with code words that resemble squeezed coherent states. However, when adding loss, we sometimes converge to codes that are invariant under rotations but are not rotation codes. Indeed, the condition for the logical $\h{Z}$ operator to be a rotation may be too restrictive. Examples of such codes include GKP codes and the numerically optimized codes plotted in Fig. \ref{fig:starcode} and in Figs. \ref{fig:phaseplot}(k) and \ref{fig:phaseplot}(l). 
 More precisely, a qudit code $\mathcal{C}$ has rotation symmetry (as opposed to being a rotation code) $N$ if its maximally mixed state $\hrho_{\mathcal{C}}=\frac{1}{d}\sum_{k=0}^{d-1}\Newketbra{k}{k}_{\mathcal{C}}$ commutes with the rotation $\h{R}_{N}=e^{\frac{2i\pi}{N}\n}$.

Alternatively, the code words can be chosen to have a defined modularity mod $N$:
\begin{claim}\label{rotationallysymmeticcodesclaim}
If a qudit code $\mathcal{C}$ has rotation symmetry $N$ and there exist integers $l_0,\ldots,l_{d-1}$ that are all different modulo $N$ s.t. $\Newmel{l_i}{\hrho_{\mathcal{C}}}{l_i}>0$ for all $i$, then we can choose different code words $\Newket{k}'_{\mathcal{C}},\,k=0,\ldots d-1$ s.t. $\Newket{k}'_{\mathcal{C}}$ is supported on $\mathrm{span}\curlbrackets{\Newket{l}|\,l=l_k (\mathrm{mod} N)}$.
\end{claim}
\begin{proof} The operators 
$\hrho_{\mathcal{C}}$ and $\h{R}_N$ commute, and both are diagonalizable. Therefore, they are simultaneously diagonalizable and we can find eigenstates of $\hrho_{\mathcal{C}}$ that are also eigenstates of $\h{R}_N$. Since the eigenvalues of $\hrho_{\mathcal{C}}$ are $\frac{1}{d},\ldots,\frac{1}{d},0,\ldots$, we obtain $d$ eigenstates $\Newket{k}'_{\mathcal{C}}$ of $\hrho_{\mathcal{C}}$ with eigenvalue $\frac{1}{d}$. Because the states $\Newket{k}'_{\mathcal{C}}$ are also eigenstates of $\h{R}_N$, they each lie in $V_{s_k} = \mathrm{span}\curlbrackets{\Newket{l}|\,l=s_k (\mathrm{mod} N)}$ for some integer $s_k$. However, since $\hrho_{\mathcal{C}}$ overlaps with $\Newket{l_i}$ for all $i=0,\ldots,d-1$ and they are all different modulo $N$, the $s_k$'s must be different. Therefore, $\Newket{k}'_{\mathcal{C}}$ can be chosen to lie in $V_{l_k}$.
\end{proof}
\subsection{\label{numeric} Numerical codes} 
\begin{figure}[h!]
  \includegraphics[width=0.45\textwidth]{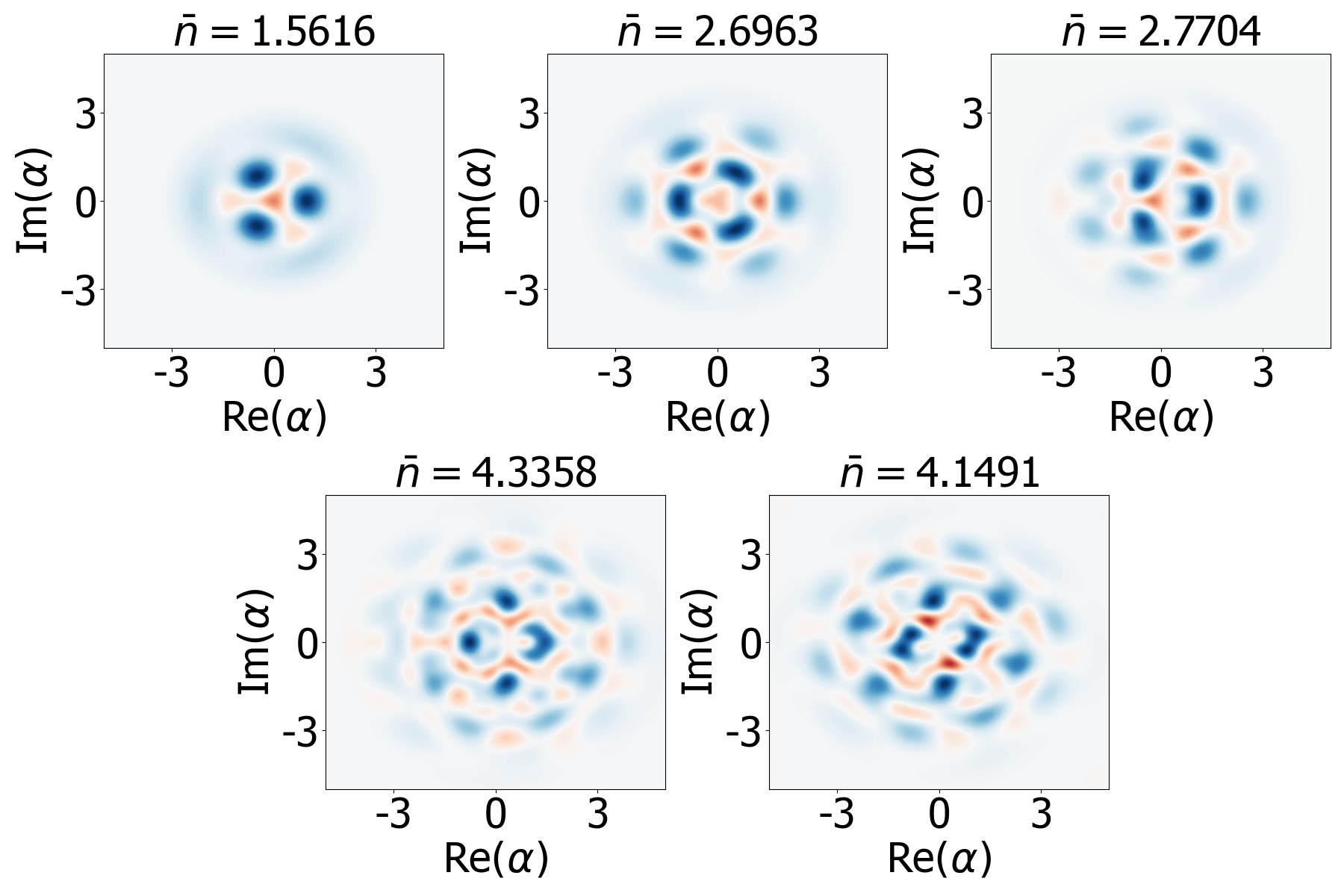}
  \caption{Wigner plots of the five numerical codes from Ref. \cite{Albert2018} and their mean energies. The top left plot depicts the $\sqrt{17}$ code.}
  \label{fig:numericcodes}
\end{figure}

Numerical codes, first introduced in Ref. \cite{Michael2016NewMode}, emerge from a different numerical optimization scheme than the one used here. For example, these codes include the $\sqrt{17}$ code
\begin{equation*}
    \begin{split}
        \Newket{0}_{\sqrt{17}}=&\frac{1}{\sqrt{6}}\brackets{\sqrt{7-\sqrt{17}}\Newket{0}+\sqrt{\sqrt{17}-1}\Newket{3}}\\
        \Newket{1}_{\sqrt{17}}=&\frac{1}{\sqrt{6}}\brackets{\sqrt{9-\sqrt{17}}\Newket{1}+\sqrt{\sqrt{17}-3}\Newket{4}}.
    \end{split}
\end{equation*}
The cost function of the optimization process is a function of coefficients of the quantum error correction matrix, which is constructed from a finite set of noise operators. In the original article \cite{Michael2016NewMode} and in Ref. \cite{Albert2018}, the noise operators were taken to be $\curlbrackets{I,\h{a},\h{a}^2}$. The codes are obtained from local minima of the cost function with a penalty on average energy. Five codes were identified in this manner. We plot the Wigner distributions of their maximally mixed states in Fig. \ref{fig:numericcodes}. Note that a different set of noise operators would result in different numerically optimal codes. In particular,  the errors are given the same weights, which is not necessarily desired when considering specific error rates. For example, if the loss rate is low, we should assign a larger weight to $I,\h{a}$. If the loss rate is high, we should assign a larger weight to $\h{a},\h{a}^2$ , and perhaps even consider higher powers of $\h{a}$. 

Since the quantum error correction matrix consists of elements of the form $\Newmel{\mu}{\cre{N}_i\ann{N}_j}{\nu}$ with $\ann{N}_i=\h{a}^i,\,i=0,1,2$ in our case, numerical codes are expected to perform well under loss (of up to two photons) and dephasing (with its first Kraus operator $\sim \n$). Indeed, there is a large parameter range for which our biconvex optimization scheme outputs  codes similar to numerical codes (e.g., Fig. \ref{fig:phaseplot}(f)).
\pagebreak
\printbibliography

\end{document}